\documentclass[12pt]{article}
\usepackage{amsfonts,amsmath,amssymb,epsf}
\usepackage{graphicx,color}
\usepackage[hypertex]{hyperref}
\usepackage{srcltx}
\newcommand{\al}{\alpha'}
\newcommand{\de}{\partial}
\newcommand{\be}{\begin{equation}}
\newcommand{\ba}{\begin{eqnarray}}
\newcommand{\ea}{\end{eqnarray}}
\newcommand{\ee}{\end{equation}}

\newcommand{\f}{\frac}
\newcommand{\s}{\sqrt}
\newcommand{\vp}{\varphi}

\newcommand{\ti}{\tilde}
\newcommand{\ap}{\alpha}

\newcommand{\ddd}{\cdot\cdot\cdot}
\newcommand{\no}{\nonumber \\}
\newcommand{\la}{\langle}
\newcommand{\lb}{\rangle}
\newcommand{\ep}{\epsilon}

\begin{document}

\begin{titlepage}
\thispagestyle{empty}

\begin{flushright}
IPMU11-0053
\end{flushright}


\begin{center}
\noindent{\large \textbf{Holographic Conductivity in Disordered Systems}}\\
\vspace{2cm}
\noindent{
Shinsei Ryu$^{a}$\footnote{e-mail:sryu@berkeley.edu},
Tadashi Takayanagi$^{b}$\footnote{e-mail:tadashi.takayanagi@ipmu.jp}, and
Tomonori Ugajin$^{b}$\footnote{e-mail:tomoki.ugajin@ipmu.jp}}

\vspace{2cm}
  {\it
 $^{a}$Department of Physics, University of California, Berkeley, CA 94720, USA\\
\vspace{1mm}
 $^{b}$Institute for the Physics and Mathematics of the Universe (IPMU), \\
 University of Tokyo, Kashiwa, Chiba 277-8582, Japan\\
 }

\vskip 2em
\end{center}

\begin{abstract}
The main purpose of this paper is to holographically study the
behavior of conductivity in 2+1 dimensional disordered systems.
We analyze probe D-brane systems in AdS/CFT with random closed string
and open string background fields. We give a prescription of
calculating the DC conductivity holographically in disordered
systems. In particular, we find an analytical formula of
the conductivity in the presence of codimension one randomness.
We also systematically study the AC conductivity
in various probe brane setups without disorder and find analogues of
Mott insulators.

\end{abstract}

\end{titlepage}

\newpage

\tableofcontents
\newpage

\section{Introduction}

The AdS/CFT correspondence
\cite{Maldacena,GKP,W,AdSR}
has been a powerful tool to investigate strongly
coupled condensed matter systems
because it claims that a strongly coupled gauge theory
is equivalent to a classical gravity theory which
lives in
higher dimensions
\cite{Ha,cmtref1,NRT,MG,Sac}.
Until now,
such applications include, for example,
quantum entanglement \cite{RT,NRT},
superconductors \cite{Gub,Hartnoll:2008vx,Ha,cmtref1},
non-relativistic systems \cite{NRCFT,Kachru:2008yh,Ka},
non-fermi liquids \cite{LMV,bottomup},
the quantum Hall effect
\cite{DKS,FLRT,BJLL}
and topological insulators \cite{RTI,HJK}.

Another important problem in realistic condensed matter
systems is the effect of quenched disorder due to impurities.
For example, when sufficiently strong,
disorder by itself
can turn
extended electron wavefunctions in a metal
into spatially localized ones
and thus triggers a metal-to-insulator transition
-- phenomenon known as Anderson localization and Anderson meta-insulator
transition
\cite{Anderson,four}
(for reviews see, e.g.,
\cite{Disorder,DisorderInteractionReview,AndersonT}).

In many realistic metal-insulator transition scenarios,
not only disorder but also electron-electron interactions
come into play, and we must study the interplay of
disorder and interactions.
The systems of most experiential interest include
phosphorus-doped silicon,
the plateau transition in the quantum Hall effect,
and
the observation of a possible metal-insulator
transition in two dimensions
\cite{MIT}.
Graphene, a two-dimensional sheet of graphite,
is another system where the combined effect of
disorder and interactions play a crucial role
\cite{graphenereview}.

However, dealing with systems with
both strong disorder and strong interactions
is to a large extent unsolved problem in condensed matter physics;
surprisingly little is understood theoretically in the
fully interacting disordered problem.
It is therefore very interesting to see how the impurity effect looks like
in the holographic approach.
This enables us to understand how strongly coupled systems behave in the presence of
randomness. An analysis of momentum relaxation in the presence of weak disorder has been done in
\cite{HaHe}. A holographic replica method for arbitrary strength of the randomness
has been formulated in \cite{Fujita:2008rs}. Impurity lattices have been holographically
constructed using D-branes in \cite{KKY}.
A functional renormalization scheme \cite{FRG} has been developed in \cite{Yaida} from
a holographic viewpoint. In spite of these developments, however, holographic calculations of transport coefficients such as conductivity have not been established yet (nevertheless see \cite{HuSh} for a progress in this direction using the
replica method). In this paper we will present a general framework to compute holographic conductivity in quenched disordered systems, based on the probe D-brane approach in AdS/CFT \cite{KK,KB}.

The probe D-brane analysis has two important advantages. One is that we can neglect the backreactions to closed string fields. The other is that we can obtain the conductivity in the presence of randomness even from classical calculations. In the bulk approach based on a random perturbation of scalar operators, the disorder only affects the conductivity at one loop or higher loops in the gravity dual \cite{Fujita:2008rs,HuSh}. Generally speaking, the holographic dual of a probe D-brane in AdS space is given by a Yang-Mills theory with quarks. The quarks are present in a flavor sector \cite{KK} or defect sector \cite{KR} and they are coupled to gluons and other matter fields which belong to the adjoint representation. As usual in this subject, we regard the charge under the $U(1)$ flavor symmetry as the electric charge of electrons, while
the non-abelian $SU(N)$ gauge symmetry of the Yang-Mills theory is interpreted as the emergent gauge symmetry in condensed matter systems. An electron is regarded as a gauge invariant operator
with the flavor charge. Notice that even though there is no dynamical $U(1)$ gauge field of photons, the emergent $SU(N)$ gauge interaction can be regarded as some analogue of the interactions due to the coulomb force and phonons. If we concentrate on the probe D-brane and forget about its backreaction to gravity sector, this corresponds to an open system of quarks interacting with
outer systems of gluons, which looks analogous to electrons in a phonon bath.
In this sense, it is not so unnatural to speculate that the strongly  correlated electrons in condensed matter systems are described by strongly coupled gauge theories with quarks. Throughout this paper we will mainly concentrate on the setups dual to $2+1$ dimensional systems, though its higher dimensional generalization can be done similarly.

In the first half of this paper, we give a systematic summary of the holographic calculation of AC and DC conductivity in general setups of probe branes without introducing the randomness. The calculation of conductivity in probe D-branes has been done first in \cite{KB}. For other earlier studies in this direction see e.g. \cite{MW,HPST,Pang,DNT,Kiritsis,Bannon}. We present several new numerical results of AC conductivity, which behave like various states of condensed matter systems from metals to insulators. In addition to the Anderson insulators, metals can be changed into insulators via another mechanism which is induced by strong interactions between electrons.
This is known as Mott insulators
\cite{Mott}.
The insulators we find in the holographic setups look analogous to Mott insulators in that the strong coupling effect is very important.

The latter half of this paper will devoted to the analysis of holographic conductivity in disordered systems. We take into account the disorder effects both from the random fluctuations of the $U(1)$ gauge field on the brane and those of the supergravity backgrounds of the probe brane. We obtain an analytical formula of the holographic DC conductivity in disordered systems for a codimension one randomness as well as perturbative results in more general cases of randomness. We will also see that depending on the distributions of the impurities we can interpolate between metallic and insulating phases.

This paper is organized as follows. In section two, we study general setups of probe D-branes and
calculate the AC and DC conductivity.
In section three, we present several explicit examples of calculations of the AC conductivity, which are regarded as holographic metals and insulators.
In section four, we study the DC conductivity in the presence of various randomness. In section five, we summarize our conclusions and discuss future problems.

\section{Holographic Conductivity in Probe D-branes}

\subsection{General Setup}

We start with a probe D$p$-brane extended in the $3+1$ dimensions
with the other $p-3$ dimensions compactified on a cycle $M_{p-3}$
in a string theory background. We will work within a supergravity approximation and
for an appropriate background with its holographic dual, this probe brane is dual to a quark-like flavor
degrees of freedom coupled with a bulk gluon sector \cite{KK}.
This may be intuitively thought to
be analogous to the electron systems (dual to the probe brane) in a phonon bath
(dual to the bulk supergravity) \cite{KB}.

The non-compact coordinate of the brane is denoted by $(t,x,y,z)$.
The supergravity background is specified by the following metric in
the Einstein frame \be
ds^2_{Ein}=R^2\left(\f{dz^2}{z^2f(z)}-\f{h(z)}{z^2}dt^2+\f{dx^2+dy^2}{z^2}\right)+ds^2_6,
\label{metricp} \ee and the string coupling constant given in terms
of $z$ dependent dilaton \be g_s=g^{(0)}_s\cdot e^{\phi(z)}. \ee For
various examples of such dilatonic solutions refer to
\cite{KRT,Kiritsis}. We shift that the value of the dilaton and
define $g^{(0)}_s$ such that it vanishes as $\phi(0)=0$ at the
boundary $z=0$ of the supergravity background (\ref{metricp}).
Notice that the six dimensional space described by the metric
$ds^2_6$ is not necessarily compact; it can include non-compact
coordinates in which the D$p$-brane is not extended. For example, a
probe D3-brane can extend in the direction of AdS$_4$ in a AdS$_5$
spacetime. On the other hand, $ds^2_6$ also includes a compact $p-3$
dimensional cycle $M_{p-3}$ on which the D$p$-brane is wrapped. The
string frame metric $G_S$ is related to the Einstein frame metric
$G_E$ via $G_{E}=e^{-\phi/2}G_{S}$. For gravity duals of the finite
temperature systems, the functions $f(z)$ and $h(z)$ have a zero at
the horizon $z=z_H$. The temperature $T$ is inversely proportional
to $z_H$ as usual. In this paper we assume that the four dimensional
spacetime spanned by $(t,x,y,z)$ is asymptotically AdS, though we
can generalize our analysis to higher dimensions without difficulty.

The DBI action of a probe D$p$-brane looks like
\footnote{
The effective action of a D-brane also includes Chern-Simons terms which express couplings to RR fields.
However, in this paper we only consider probe brane configurations where we can neglect such terms.
The Chern-Simons terms become
important when we are interested in topological properties
as observed in the quantum Hall effect
\cite{DKS,FLRT,BJLL}
and topological insulators \cite{RTI,HJK}.
}
\ba
&& S^{Dp}_{DBI}=-\tau_p \int (d\xi)^{p+1} e^{-\vp(z)}\s{-\det(G_S+F)}, \\
&& =-\f{T_p}{(2\pi\al)^2}\int dtdxdydz~ e^{-\vp(z)}\s{-\det(G_Ee^{\phi/2}+F)},
\ea
where $\xi$ denotes the $p+1$ coordinates in which the brane extends.
The field strength of the $U(1)$ gauge field on the brane is denoted by $F=dA$. The constant
$\tau_p=\f{1}{g^{(0)}_s (2\pi)^p(\al)^{(p+1)/2}}$ is the standard tension of a
D$p$-brane with the string coupling constant evaluated at the boundary $z=0$.
$\vp(z)$ depends both on the dilaton and the volume of $M_{p-3}$. More explicitly, we have
\be
T_p\cdot  e^{-\vp(z)}=(2\pi\al)^2\cdot\tau_p\cdot V_{p-3}(z)\cdot e^{-\phi(z)},
\ee
 where $V_{p-3}(z)$ is the $z$ dependent volume of $M_{p-3}$ in string frame. We fix the constant part of
 $\vp(z)$ by the condition $\vp(0)=0$. In particular, if $V_{p-3}$ does not depend on
$z$, then we simply find $\vp(z)=\phi(z)$. Notice that $T_p$ is normalized such that it is dimensionless.

Explicitly we can evaluate the DBI action as follows
\be
S_{DBI}=-\f{R^2T_p}{(2\pi\al)^2}\int \f{dtdxdydz}{z^4} e^{-\vp+\phi}\s{\f{hR^4}{f}-e^{-\phi}z^4F_{tz}^2-\f{z^4e^{-\phi}}{f}F_{tx}^2
+z^4he^{-\phi}F_{zx}^2}. \label{dbis}
\ee
In this paper, we will work in the gauge $A_z=0$,

In order to have a holographic dual with non-zero charged density, we need a solution with the electric
flux $F_{tz}$. This is given by
\be
A_t(z)=\mu+\rho\int^z_0 dy\s{\f{h(y)}{f(y)F(y)}}, \label{gpa}
\ee
where we defined
\be
F(z)=e^{-2\vp(z)}+\f{\rho^2}{R^4}z^4e^{-\phi(z)}. \label{defp}
\ee
Notice that in our normalization, i.e. $\phi(0)=\vp(0)=0$, we always find $F(0)=1$ at the boundary.

According to the standard bulk to boundary relation in gravity duals, in our model (\ref{dbis})
the charge density and current are given by
\ba
&& j_t=-\f{\delta S_{DBI}}{\delta A_t}=T_p\lim_{z\to 0}\left[\s{\f{f(z)F(z)}{h(z)}}F_{zt}(z)\right]=T_{p}F_{zt}(0)
\label{chd},\\
&&  j_x=-\f{\delta S_{DBI}}{\delta A_x}=-T_p \lim_{z\to 0}\left[\s{\f{f(z)F(z)}{h(z)}}F_{zx}(z)\right]=-T_pF_{zx}(0).    \label{bbrel}
\ea
Here we rescaled the gauge field as $A_\mu\to 2\pi\al A_{\mu}$ so that the unit electric charge of the F-string is normalized to be one. This also replaces $\rho^2/R^{4}$ in (\ref{defp}) with
$(2\pi\al)^2\rho^2/R^4$ and thus
in general AdS/CFT setups, the function $F(z)$ takes the form
\be
F(z)=\s{a_1(\lambda)+a_2(\lambda)\cdot\rho^2z^4}, \label{tempdepc}
\ee
where $a_1$ and $a_2$ are certain functions of the 't Hooft coupling $\lambda$, which depend on the details of
setups.
For the electric solution (\ref{gpa}), we find that $\mu$ corresponds to the chemical potential and $\rho$ is proportional to the charge density via the relation $j_t=T_p\rho$. In this paper, we will always set
$2\pi\al=1$ just for a presentational simplification of our results.

\subsection{Calculating Holographic Conductivity}

By expanding the action around this solution (\ref{gpa}) at the quadratic order about $A_x$, we obtain
\be
S_{DBI}\simeq T_p\int dtdxdydz\s{\f{f(z)F(z)}{h(z)}}\left(\f{F_{tx}^2}{f(z)}-h(z) F_{xz}^2\right).
\ee

The equation of motion for $A_x$ with the frequency $\omega$ i.e. $A_x \propto e^{-i\omega t}$ reads
\be
A''_x+\left(\f{\de_z\s{h(z)f(z)}}{\s{h(z)f(z)}}+g(z)\right)A'_x+\f{\omega^2}{h(z)f(z)}A_x=0, \label{axeqone}
\ee
where we defined
\be
g(z)=\f{\de_z F(z)}{2F(z)}.\label{defg}
\ee

It is useful to employ a new radial coordinate $w$ instead of $z$ defined by
\be
w(z)=\int^z_0 \f{dy}{\s{h(y)f(y)}}. \label{wz}
\ee
The boundary $z=0$ corresponds to $w=0$.
In all of our examples, $w$ runs from $0$ to $\infty$ and the
black brane horizon $z=z_H$ corresponds to $w=\infty$.

At the same time, we rescale the gauge field $A_x$ into a new field $\Psi(w)$
\be
A_x(w)=F(z)^{-1/4}\cdot \Psi(w).
\ee

After this coordinate change $z=z(w)$ and rescaling of $A_x$, the Maxwell equation (\ref{axeqone})
takes the form of the Schrodinger equation
\be
-\Psi''(w)+V(w)\Psi(w)=\omega^2\Psi(w).\label{Sch}
\ee
The effective potential is given by
\ba
V(w)&=&\f{1}{4}f(z)h(z)g(z)^2+\f{1}{2}\de_w \left[g(z)\s{h(z)f(z)}\right]\no
&=& \f{1}{4}f(z)h(z)g(z)^2+\f{1}{2}\s{h(z)f(z)}\de_z \left[g(z)\s{h(z)f(z)}\right]. \label{effpot}
\ea

As pointed out in \cite{Azeyanagi:2009pr}, we can relate the holographic calculation of conductivity to
the following scattering problem in the Schrodinger problem.
For this purpose, it is convenient to extend the range of the coordinate $w$ to $w<0$ assuming
$V(w)=0$ there. Next we consider the following boundary value problem. First we impose the in-going boundary condition at the horizon $w=\infty$ \cite{SS}
\be
\Psi(w)\to T(\omega)\cdot e^{i\omega w}\ \ \ (w\to \infty),  \label{holbc}
\ee
where we implicitly
use the fact that the potential $V(w)$ gets vanishing in this limit $w=\infty$
in all examples we are interested in. Then we solve (\ref{Sch}) with this boundary condition toward the boundary
$w=0$ to find the reflection coefficient $R(\omega)$  defined by
\be
\Psi(w)\to e^{i\omega w}+R(\omega)e^{-i\omega w}\ \ \ \ (w\to 0).
\ee
In this way, we can uniquely obtain the reflection
coefficient $R(\omega)$ and the transmission coefficient $T(\omega)$, which satisfy
$|R(\omega)|^2+|T(\omega)|^2=1.$ At the same time, this allows us to determine the holographic conductivity uniquely as follows \be
\sigma(\omega)=\f{j_x}{E_x}=\f{T_p\cdot \de_w A_x(0)}{i\omega A_x(0)}=T_p\cdot \f{\de_w \Psi(0)}{i\omega \Psi(0)}
=T_p\cdot \f{1-R(\omega)}{1+R(\omega)}.\label{cond}
\ee

To evaluate the AC conductivity $\sigma(\omega)$
explicitly, we need to solve (\ref{Sch}) numerically in general, as we will do in several examples later.
However, it will be helpful to look at the simplest example where we find an exact expression of $\sigma(\omega)$.  Consider the case where $g(z)$ vanishes and for example, this is realized if the charged density is vanishing $\rho=0$ and  $\vp(z)$ is constant. Even apart from our probe D-brane approximation, this corresponds to the case of the minimal Einstein-Maxwell theory with zero charge density. In such cases, the potential (\ref{effpot}) clearly vanishes and thus there is no reflection $R(\omega)=0$.
Therefore we find from (\ref{cond}) that the conductivity is independent of $\omega$
\be
\sigma(\omega)=T_p ~.
\ee

Even for general examples, since the effect of potential $V(w)$ gets smaller as we increase the energy $\omega^2$, we always find that in the high frequency limit $\omega/T\to\infty$
\be
\lim_{\omega\to \infty}\sigma(\omega)=T_p ~. \label{proplo}
\ee

In the opposite case, where the potential $V(w)$ gives so large barrier that makes the tunneling effect impossible, we will find that the reflection coefficient $R(\omega)$ is purely a phase
factor $e^{i\delta(\omega)}$ for small values of $\omega$. Then the real part of the conductivity $\mathrm{Re}\, \sigma(\omega)$ is
vanishing.  Such systems correspond to insulators because the DC conductivity clearly vanishes, while the others
in our probe setup can be regarded as metals.

Another property which can be found analytically in generic cases is the sum rule. As we will explain in the appendix \ref{setion:sumrule}, if there is no delta functional Drude peak, the AC conductivity in our setup always satisfies the simple sum rule (see also \cite{Hartnoll:2008vx})
\be
\int^\infty_0 d\omega \left(\mbox{Re}\, \sigma(\omega)-T_p \right)=0.
\ee

\subsection{DC Conductivity}

Even for general systems, we can calculate the DC conductivity $\sigma_{DC}=\sigma(0)$ in an analytical way.
This has been first calculated in \cite{KB} for several examples of probe D-branes in a different way. Here we derive the expression of $\sigma_{DC}$ as the $\omega\to 0$ limit of the AC conductivity $\sigma(\omega)$.

For this, it is useful to define a new function $X$ by
\be
X(w)\equiv \f{\de_w A_x(w)}{A_x(w)}=\f{\de_w\Psi}{\Psi}-\f{1}{2}g(z)\s{h(z)f(z)}. \label{relax}
\ee
We often write equations using both the coordinate $z$ and $w$, as they are related by (\ref{wz}).
The equation of motion (\ref{axeqone}) is rewritten to be
\be
\de_w X=-gX\s{hf}-X^2-\omega^2, \label{xeom}
\ee
and the boundary condition at the horizon (\ref{holbc}) is simply expressed as follows
\be
X(\infty)=i\omega.  \label{bcxh}
\ee
Notice that (\ref{bcxh}) follows from (\ref{holbc}) by assuming lim$_{w\to\infty}g\s{fh}=0$, which
is essentially equivalent to the previous assumption $\lim_{w\to\infty}V(w)=0$.

The conductivity $\sigma(\omega)$ is found from the value of $X$ at the boundary $w=0$ via
\be
\sigma(\omega)=T_p\cdot\f{X(0)}{i\omega}. \label{holxcond}
\ee
In the $\omega\to\infty$ limit, we find that the solution to (\ref{xeom}) approaches to
$X(w)=i\omega$ and we reproduce the behavior (\ref{proplo}).
Below we would like to find the behavior of $X(\omega)$ in the opposite limit $\omega\to 0$ to obtain the
DC conductivity.

We can see from (\ref{xeom}) and (\ref{bcxh}) that the real part
$\mathrm{Re}\, \sigma(\omega)$ and the imaginary
part
$\mathrm{Im}\, \sigma(\omega)$ are even and odd function of $\omega$, respectively. They are also related by the
Kramers-Kronig relation
\be
\mbox{Im}\, \sigma(\omega)=\f{1}{\pi}\cdot P\int^{\infty}_{-\infty} d\omega'~
\f{\mbox{Re}\, \sigma(\omega')-T_p}{\omega-\omega'},
\label{KKR}
\ee
and thus if we know one of them, we can reproduce the other.

Let us first examine the
behavior of the imaginary part in the $\omega\to 0$ limit. The non-vanishing $X$ in the $\omega\to 0$ limit
means a pole at $\omega=0$ as follows from (\ref{holxcond}). This pole leads to a delta-functional
peak in
$\mathrm{Re}\, \sigma(\omega)$ as it can be directly confirmed by using (\ref{KKR}). At finite temperature (or finite value of $z_H$), we can show that this pole
does not appear in our probe systems because we can find the simple regular solution $X=0$ to (\ref{xeom}) and (\ref{bcxh}) at $\omega=0$. This is natural as the phonon bath will introduce the dissipation and make the
DC conductivity finite. On the other hand, the solution $X=0$ gets singular at zero temperature limit $z_H\to \infty$ unless
 \be
\int^\infty_0 dw g(w)\s{h(z)f(z)}= \int^{z_H}_0 dz g(z)<\infty. \label{finitec}
 \ee

Next we turn to the real part $\mathrm{Re}\, \sigma(\omega)$ in the $\omega\to 0$ limit. Since $X$ is order
$O(\omega)$, we can neglect
$O(\omega^2)$ terms in (\ref{xeom}). This leads to
\be
X(w)=i\omega\cdot e^{\int^\infty_w dwg(w)\s{h(z)f(z)}}=i\omega\s{F(z_H)},
\ee
where we employed (\ref{defg}) and our normalization $F(0)=1$.
In this way, we obtain the simple expression of the DC conductivity $\sigma_{DC}$
 \ba
&& \sigma_{DC}\equiv \mbox{Re}\ \sigma(0) \no
&&\ \ \  \ \ \  =
 T_p\s{F(z_H)}=T_p\s{e^{-2\vp(z_H)}+\f{\rho^2}{R^4}z_H^4e^{-\phi(z_H)}},
\label{condx}
 \ea
 which is finite and positive. If we remember that we set $2\pi\al=1$,
 we find that the conductivity (\ref{condx}) is dimensionless as is typical
 in 2+1 dimensional systems. For non-dilatonic backgrounds, by using (\ref{tempdepc}),
 the holographic DC conductivity takes the form
 \be
 \sigma_{DC}=\s{b_1(\lambda)+b_2(\lambda)\f{\rho^2}{T^4}}, \label{DCcondp}
 \ee
where $b_1(\lambda)$ and $b_2(\lambda)$ are certain model-depending
dimensionless functions of the 't Hooft coupling $\lambda$.
$\sigma_{DC}$ gets divergent at zero temperature
for a non-vanishing charge density $\rho$. This is expected since
there are no active gluons (or `phonons') at zero temperature and there is no dissipation.

It might also be useful to compare (\ref{DCcondp})
qualitatively
with standard results in electron systems, though in our holographic case there is no clear quasi particle picture. The standard Drude
formula is given by $\sigma_{DC}=\f{ne^2\tau}{m}$, where $n$, $e$, $\tau$ and $m$ denote
the electron density, charge, mean free time and mass. In our setup we can estimate
$e=1$, $n\sim \rho$, $\tau\sim 1/T$ and $m\sim T$. This indeed explains the behavior
$\sigma_{DC}\sim \f{\rho}{T^2}$ of (\ref{DCcondp}) when $\rho$ is very large.

Finally, it is intriguing to explore possibilities of charged insulator systems in our probe brane setups.
As is clear from (\ref{condx}), insulators are possible only if $e^{-2\vp(z_H)}=\rho^2z_H^4e^{-\phi(z_H)}=0$.
At zero temperature $z_H\to \infty$, this requires the infinitely large value of the dilaton in the IR limit of the supergravity background. In the dual gauge theory language, the IR theory gets infinitely strongly coupled. In this sense, we may think that this is analogous to Mott insulators, even though its gravity dual gets singular.
In general, it is evident from (\ref{condx}) that the conductivity is reduced (or increases) if the
dilaton gets large (or small) in the IR. We will present explicit results of AC conductivity in various systems from metals to insulators in the next section. Those readers who are mainly interested in holographic conductivities in disordered systems
may jump to section \ref{section:disorder}.

\section{AC Conductivity: From Metals to Insulators}\label{setion:examples}

\subsection{Charged Probe Brane in AdS: Holographic Metal}
\begin{figure}[htbp]
\begin{center}
    \includegraphics[width=5cm]{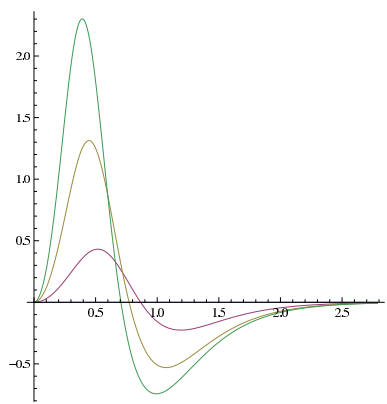}
\end{center}
\caption{The effective potential $V(w)$ for the charge probe D-brane in AdS. It corresponds to $\rho=0(\mbox{blue}),1(\mbox{red}),2(\mbox{yellow}),3(\mbox{green})$.}
\label{potentialprobe}
\end{figure}

\begin{figure}[htbp]
\begin{center}
    \includegraphics[width=6cm]{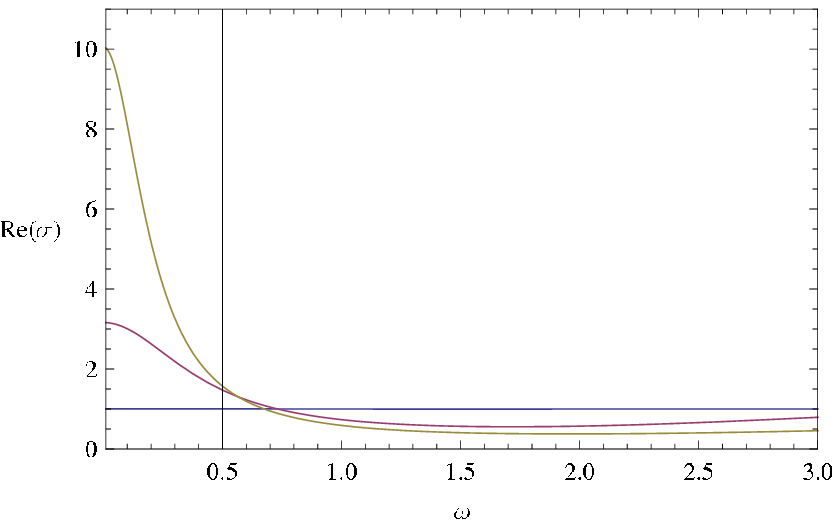}
        \hspace*{0.5cm}
    \includegraphics[width=6cm]{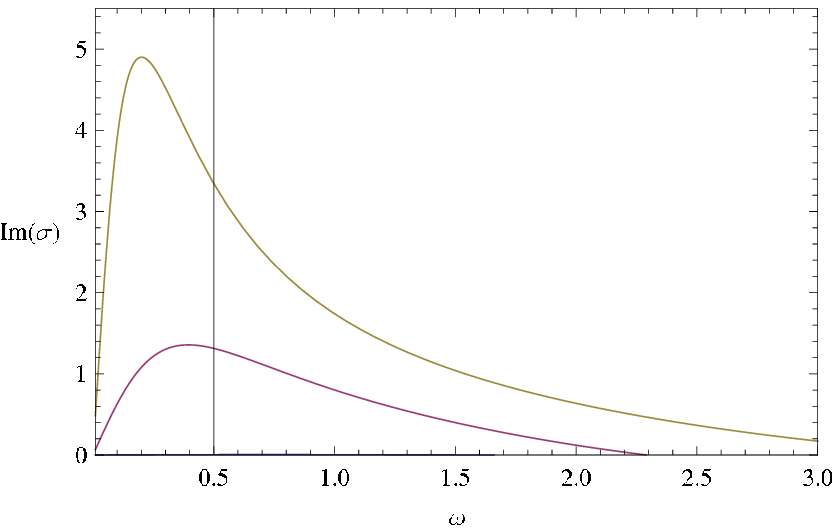}
\end{center}
\caption{The real and imaginary part of conductivity from charged probe brane in AdS. We took $\rho=0.1(\mbox{blue}),3(\mbox{red}),10(\mbox{yellow})$.}
\label{probeclean}
\end{figure}

The most basic example of our probe D-brane setup is the charged probe D-brane in AdS
Schwarzschild black branes. Though this has already been discussed in several earlier works
e.g. \cite{KB,HPST,DNT}, we first study this system from our viewpoint because this is the most basic example.
Then we can assume that the metric (\ref{metricp}) is given by the codimension one subspace of $AdS_5$ black brane \be
f(z)=h(z)=1-\left(\f{z}{z_H}\right)^4,
\ee
and that the dilaton and the compactified volume take constant values, which allows us to simply set
$\phi(z)=\vp(z)=0$. The functions $F(z)$ and $g(z)$ are
\be
F(z)=1+\f{\rho^2}{R^4}z^4,\ \ \ \ g(z)=\f{2\rho^2z^3}{R^4+\rho^2z^4}.  \label{FGads}
\ee
The effective potential $V(w)$ defined by (\ref{effpot}) is plotted in Fig.\ref{potentialprobe}.
Here we chose and $R=z_H=1$. As is clear from its profile, there is a positive peak followed negative peak as $w$ gets larger.
The former peak will reduce the conductivity at slightly high energy as it gives larger reflection, while the latter
enhances the tunneling in the low energy.
This expectation is indeed reflected in the plot of the
(real part of) AC conductivity shown in Fig.\ref{probeclean}. We have set $T_p=1$.
We can find a finite Drude peak in $\mathrm{Re}\, \sigma(\omega)$ and it eventually approaches to a constant
$\mathrm{Re}\, \sigma(\omega)=1$ in the large $\omega$ limit, which is peculiar to the $2+1$ dimensional critical systems as we also explained in (\ref{proplo}).
In this probe system, the gluon sector plays a role of
phonons and this leads to the dissipations. Thus we do not get the delta-functional Drude peak
but get the smooth finite Drude peak at non-zero temperature. In the zero temperature limit, it changes into a delta-functional peak as we can see from the fact that the value of $F(z_H)$, which is proportional to the DC conductivity (\ref{condx}), gets divergent in the limit $z_H\to\infty$.

\subsection{Probe Brane vs. Bulk Approach in Holographic Metals and Superconductors}

Before we go on, it is useful to remind us of holographic calculations of conductivity
in bulk gravity duals such as the ones in Einstein-Maxwell theory without introducing
any probe D-branes. We would like to make their difference clearer in spite of their similar
looking.

\begin{figure}[htbp]
\begin{center}
    \includegraphics[width=6cm]{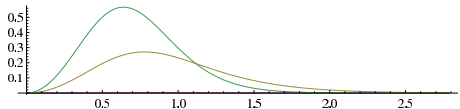}
\end{center}
\caption{The effective potential $V(w)$ for the charge AdS black brane for various charges
$Q=0,0.1,1,\s{3}$.}
\label{potentialbulk}
\end{figure}

First, it is clear that a charged AdS black brane in Einstein-Maxwell theory is dual to a metallic system. In this example, if we plot the AC conductivity, we find a delta-functional Drude peak
even at non-zero temperature. This unusual behavior is because the system does not coupled to any other sectors as opposed to the previous probe D-brane example and thus there are no external objects analogous to phonons. This fact is manifest in the equation of motion of the gauge field, which is obtained after we combine the Einstein equation with Maxwell equation as calculated in \cite{Ha}
\be
A''_x+\left(\f{f'(z)}{f(z)}\right)A_x'+\left(\f{\omega^2}{f(z)^2}-\f{Q^2z^2}{f(z)}\right) A_x=0, \label{maxx}
\ee
where we set $f(z)=1-(1+Q^2)z^3+Q^2z^4$. The charge of the AdS black brane is denoted by $Q$.
The important point is that the effect of non-zero charge enters via the coefficient in front of
$A_x$. This makes a sharp contrast with the probe D-brane case where the charge effect appears in
the coefficient of $A'_x$ term as in (\ref{axeqone}).
In this bulk gravity dual case, we have  instead of (\ref{xeom})
\be
\de_wX=-X^2-\omega^2-Q^2z^2f(z).
\ee
Thus $X$ gets non-vanishing at the boundary $z=0$, leading to a pole of
Im$\sigma(\omega)$ at $\omega=0$, leading to a delta functional Drude peak via the
Kramers-Kronig relation. The Schrodinger potential for the equation (\ref{maxx}) is given by
\be
V(w)=Q^2 z^2f(z). \label{bulkpot}
\ee
As the $Q$ increases, the potential barrier grows. Therefore the reflection increases
and thus Re $\sigma(\omega)$ is reduced for non-zero small $\omega$. The explicit form of the potential $V(w)$ is
plotted in Fig.\ref{potentialbulk}.

One may think that this delta functional Drude peak is very similar to
that of holographic superconductors \cite{Hartnoll:2008vx,Ha,cmtref1}
as superconductors are characterized by an
infinitely large DC conductivity. Indeed, in the bulk gravity dual, the calculation
of conductivity for holographic superconductors goes very similarly. Basically,
we just need to replace $Q^2$ with $Q^2+|\Phi|^2$ in (\ref{maxx}) and (\ref{bulkpot}),
where $\Phi$ is a complex scalar field whose condense triggers the superconductor
phase transition. Therefore the effect of $\Phi$ condensation is very similar to that of
non-zero charge $Q$.

However, once we promote them to probe D-brane setups \cite{Ammon:2008fc}, we
will notice the crucial difference between holographic superconductors and charged AdS black branes.
At finite temperature, the former still has a delta functional Drude peak as the effect of scalar field always come in $A_x$ term, while the latter leads to a finite Drude peak as the effect only appears in the $A'_x$ term. In this way, there is a sharp distinction between the
superconductors and metals in the probe D-brane approach.

\subsection{Holographic Metals with Pseudo Gaps}

It is very interesting to understand what will happen if we modify the
supergravity background of a charged probe brane. One important class of modifications is to assume a non-trivial profile of the dilaton $\phi$. Another will be to consider different metrics than
the AdS, such as the Lifshitz backgrounds \cite{Kachru:2008yh}. Since there has already been considerable works\footnote{For the analysis of holographic conductivity in Lifshitz backgrounds
refer to e.g. \cite{Ka,Pang,Kiritsis,Bannon}.} in the latter, we will focus on the former below.

Basically, there are two possibilities of the behavior of dilaton at zero temperature: the dilaton gets smaller or larger in the IR limit $w\to\infty$.
In this subsection we will study the former, while the latter will be discussed in the next subsection.

Here we will study the type IIB
supergravity solution found in \cite{Azeyanagi:2009pr} as an example of such a dilatonic
background. The metric of this black brane solution reads
in the Einstein frame
\ba
&& ds^2=R^2\left[z^{-2}(-F(r)dt^2+dx^2+dy^2)+z^{-4/3}d\chi^2+\f{dz^2}{z^2F(z)}\right]+\ti{R}^2ds^2_{X_5}, \no
&& F(r)=1-M\cdot z^{11/3}, \label{ALTIIB}
\ea
where $X_5$ is an arbitrary Einstein manifolds. $M$ is a mass parameter of the black brane which is
proportional to the temperature. On the other hand, the dilaton behaves
$e^{\phi(z)}\propto z^{-2/3}$ and goes to zero in the IR limit $z\to \infty$
as we promised. This solution is supported by RR 5-form and 1-form fluxes and can be regarded as a gravity dual of (non-supersymmetric) D3-D7 system. The holographic dual gauge theory lives in the $3+1$ dimension spanned by $(t,x,y,\chi)$ and from this viewpoint, the $\chi$ coordinate has an anisotropic scaling property. In this sense, this background (\ref{ALTIIB}) can be regarded as a gravity dual of space-like Lifshitz fixed point. However, for our purpose, a probe D3-brane is assumed to extend in $(t,x,y,r)$ direction in $X_5$
and therefore the probe brane world-volume looks like a AdS space or its black brane, rather than a
Lifshitz background. Notice that in this setup we have $\phi(z)=\vp(z)$ because the radius of $S^5$ denoted by
$\ti{R}$ does not depend on $z$.

As shown in \cite{Azeyanagi:2009pr}, we can construct an interpolating solution which starts from the conventional AdS$_5\times$S$^5$ solution at the boundary $z=0$ and approaches the dilatonic
solution (\ref{ALTIIB}) in the IR limit $z\to \infty$. Since this solution is highly complicated, below we will replace it with the simple profile of the dilaton
\be
e^{\phi(z)}=e^{\vp(z)}=\f{1}{(1+z^2)^{1/3}},\label{ALTdilaton}
\ee
 with the metric (\ref{ALTIIB}) unchanged. Since the relevant physics is essentially hidden in
the scaling metric (\ref{ALTIIB}), this approximation is enough for our purpose.

\begin{figure}[htbp]
\begin{center}
    \includegraphics[width=5cm]{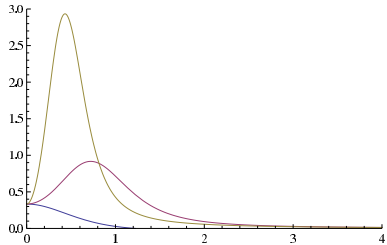}
\end{center}
\caption{The effective potential $V(w)$ of charged probe brane in the dilatonic background  (\ref{ALTIIB}) at zero temperature
for $\rho=0$ (blue), $\rho=1$ (red) and $\rho=3$ (yellow).}
\label{potentialzeroALT}
\end{figure}

\begin{figure}[htbp]
\begin{center}
    \includegraphics[width=5cm]{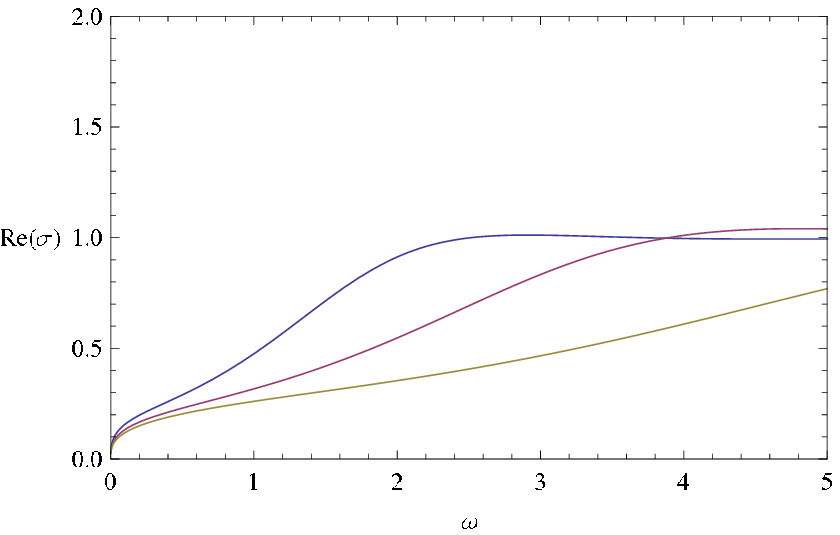}
\hspace*{0.5cm}
    \includegraphics[width=5cm]{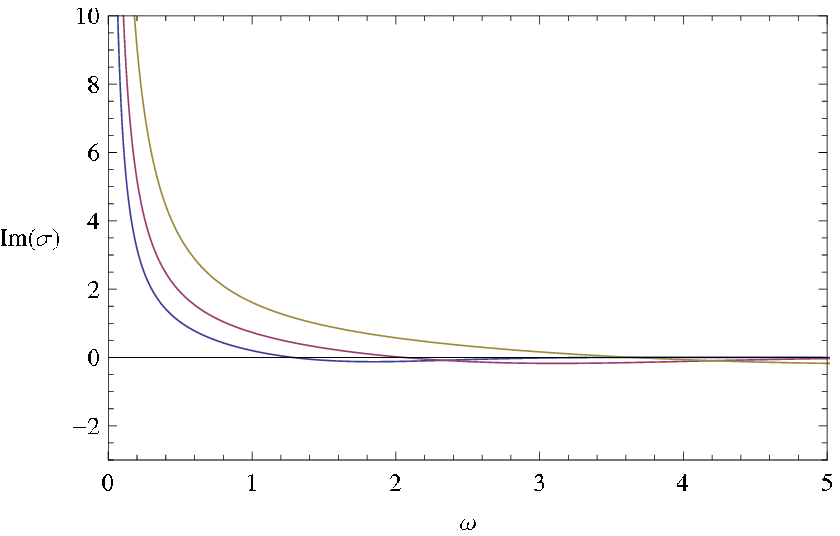}
\end{center}
\caption{The plot of real and imaginary part of the conductivity for the probe brane in the dilatonic background  (\ref{ALTIIB}) at zero temperature
for $\rho=1$ (blue), $\rho=3$ (red) and $\rho=10$ (yellow). }
\label{condzeroALT}
\end{figure}

First we start with calculating the holographic conductivity at zero
temperature $M=0$. The effective potential $V(z)$ behaves like \be
V(z)\to \f{1}{3}\ \  (z\to 0),\ \ \ \ \ V(z)\to \f{7}{36 z^2} \ \
(z\to\infty),\label{potalta} \ee and its explicit form of the
potential is plotted in Fig.\ref{potentialzeroALT}. After solving
the Schrodinger problem, the resulting conductivity is shown in
Fig.\ref{condzeroALT}. There is a delta-functional Drude peak in Re
$\sigma(\omega)$ hidden at $\omega=0$ as is clear from the behavior
of Im $\sigma(\omega)$ at $\omega=0$ and also from the fact that
$F(z_H)\to \infty$  in the limit $z_H\to\infty$. Also we can confirm
for small $\omega$ \be \mbox{Re}~\sigma(\omega)\propto \omega^{1/3}\
\ \left(+\delta(\omega)\ \mbox{term}\right). \ee The reason why this
vanishes in the limit $\omega\to0$ is that the potential $V(z)\sim
z^{-2}$ for large $z$ (\ref{potalta}) does not allow tunneling at
zero energy\footnote{Such a power behavior of the AC conductivity
has been already known in various calculations of holographic
conductivity \cite{Ka,Pang,Kiritsis} with or without probe
approximations. Our example here can be regarded as a particular
example of these classes and what we showed here is that we can
embed this into type IIB string theory.}. In this sense, the system
dual to the D3-brane in our dilatonic background behaves like a
metal with a pseudo gap.

\begin{figure}[htbp]
\begin{center}
    \includegraphics[width=4cm]{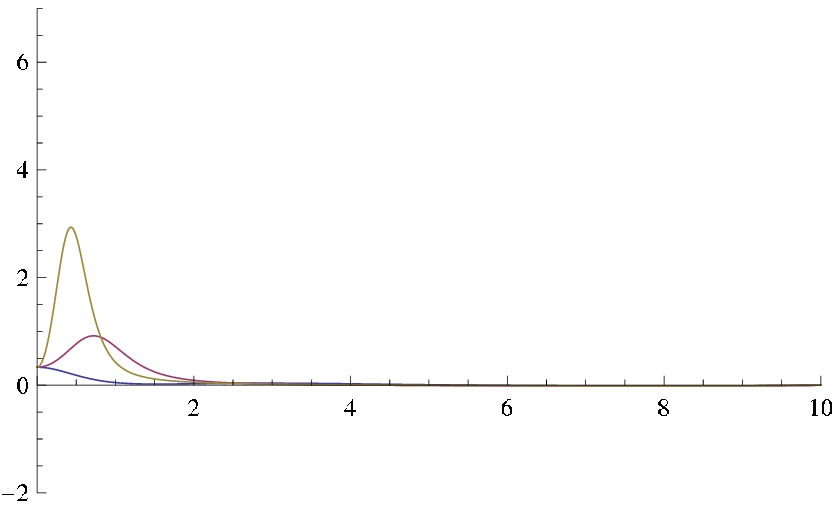}
\hspace*{0.5cm}
\includegraphics[width=4cm]{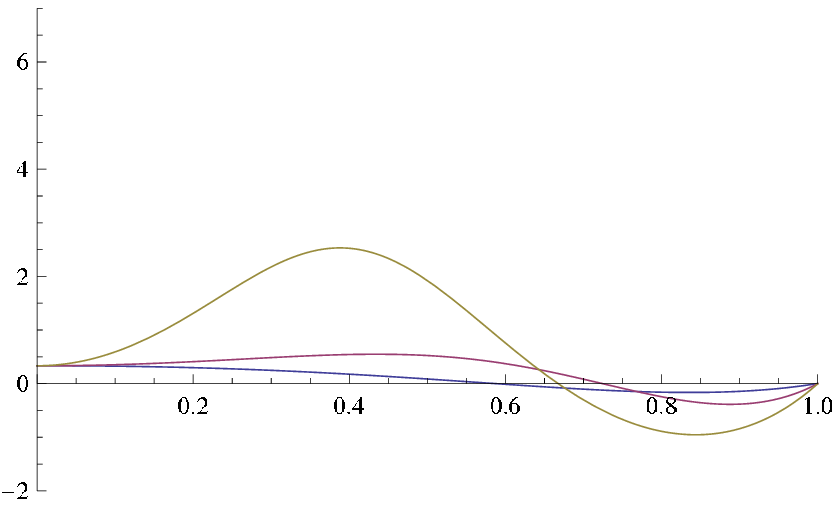}
\hspace*{0.5cm}
    \includegraphics[width=4cm]{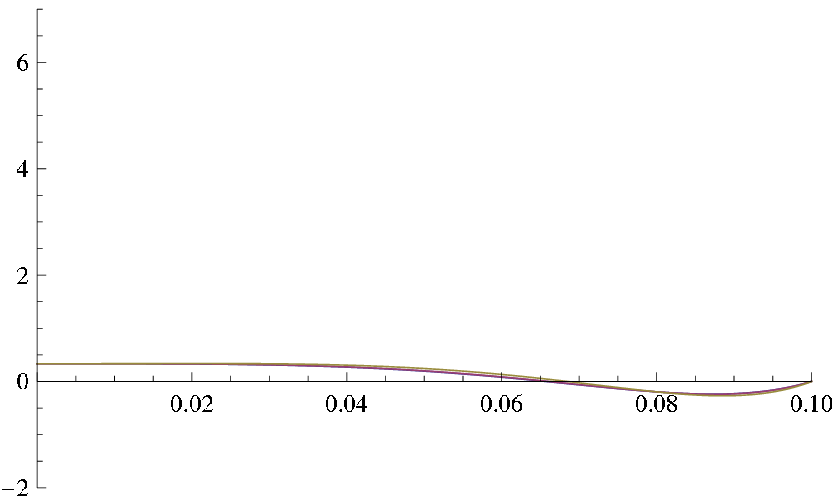}
\end{center}
\caption{The effective potential $V(z)$ for the finite temperature dilatonic solution (\ref{ALTIIB}).
The temperature is given by $M=0.1~(\mbox{left}),~M=1~(\mbox{middle}),~M=10~(\mbox{right})$. The charge density is $\rho=0.1$ (blue), $\rho=1$ (red) and $\rho=3$ (yellow).}
\label{potentialALT}
\end{figure}

\begin{figure}[htbp]
\begin{center}
    \includegraphics[width=5cm]{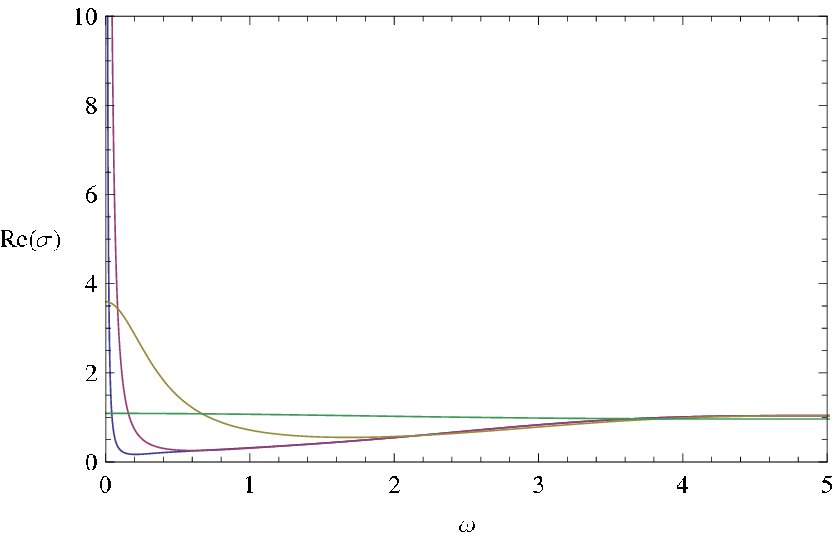}
\hspace*{0.5cm}
    \includegraphics[width=5cm]{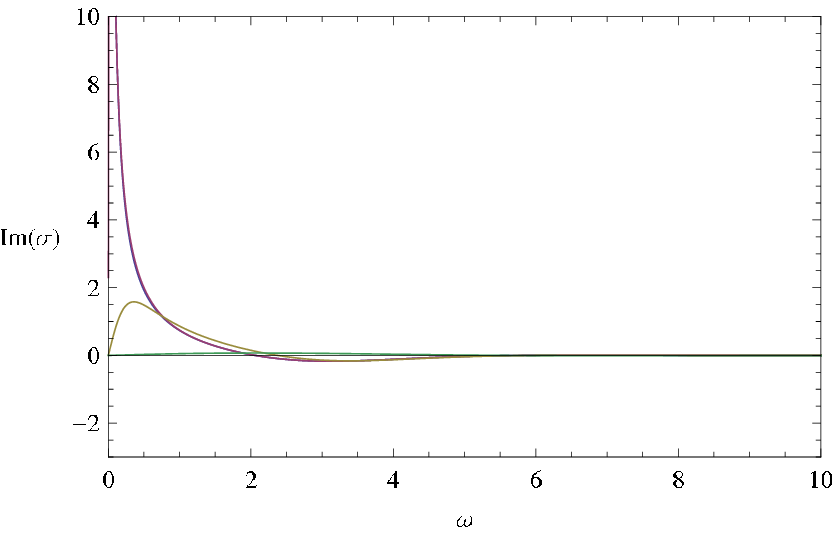}
\end{center}
\caption{The plot of real and imaginary part of the conductivity for the probe brane in the finite temperature dilatonic solution (\ref{ALTIIB}). ($M=0.1$ (blue), $M=0.3$ (red), $M=1$ (yellow), $M=3$ (green). All have the charge density $\rho=3$.}
\label{condALT}
\end{figure}

At finite temperature, corresponding to positive values of $M$ in (\ref{ALTIIB}),
the effective potential behaves like
\be
V(0)=\f{1}{3},\ \ \ \ \ \ \ V(w)\to O(e^{-w})\ \  (w\to\infty).
\ee
The explicit form of the potential is plotted in Fig.\ref{potentialALT}. The AC conductivity
at finite temperature is shown in Fig.\ref{condALT}.

As is clear from this potential, as the temperature increases, the sharp potential wall at small $z$ spreads and a negative potential region develops in the near horizon region $z>>1$.
Accordingly, the delta functional Drude peak at zero temperature becomes broad and
the pseudo gap gradually disappears as the temperature increases.
The high temperature behavior is similar to the AdS BH (i.e. Fig.\ref{potentialprobe}).

\subsection{Towards Holographic Mott Insulators}

As final examples, we would like to consider possibilities of
constructing gravity duals of insulators at non-zero charge
density\footnote{There are earilar works on holographic insulators.
At vanishing charge density, one clear example of holographic
insulators is the AdS soliton \cite{Nishioka:2009zj}. In the charged
black hole background, it has been shown that interactions can make
fermions gapped in \cite{ELP}. Also the system whose AC conductivity
shows the power law behavior (we called this a metal with a pseudo
gap) can be regarded as a Mott insulator if we can get rid of the
delta functional Drude peak \cite{Kiritsis,MCG}.}. For this, we need
to realize the vanishing DC conductivity at zero temperature. This
is possible in our probe D-brane setup only if the dilaton gets
infinitely large in the IR limit $z\to \infty$ as we already
mentioned. Even though in such an example we need to worry about the
singular behavior due to the strongly coupled limit, we will ignore
this issue and proceed to the calculations of conductivity assuming
that an appropriate string duality will make the background
non-singular.

\begin{figure}[htbp]
\begin{center}
    \includegraphics[width=4.5cm]{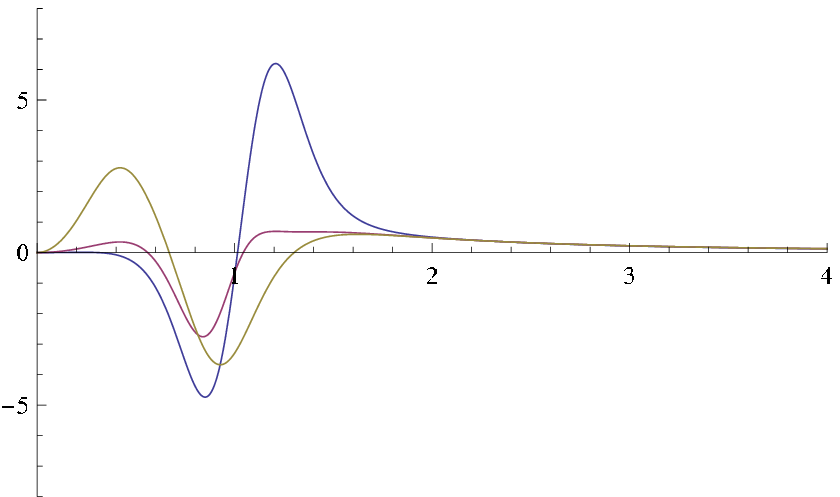}
\includegraphics[width=4.5cm]{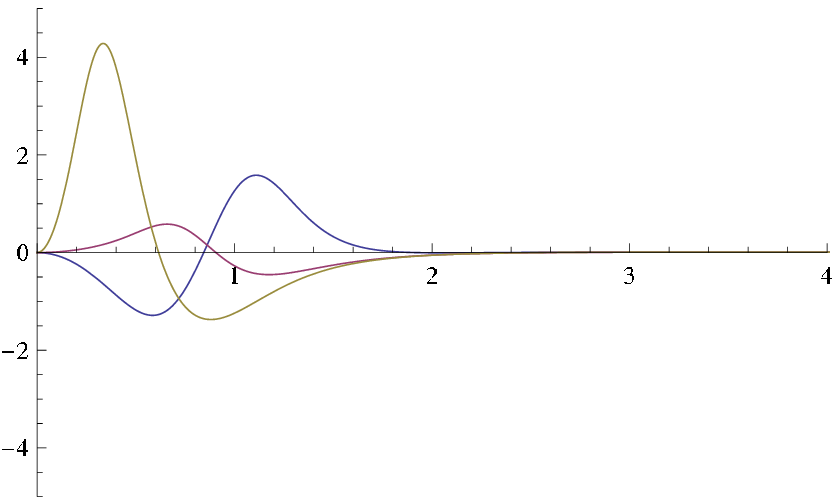}
\includegraphics[width=4.5cm]{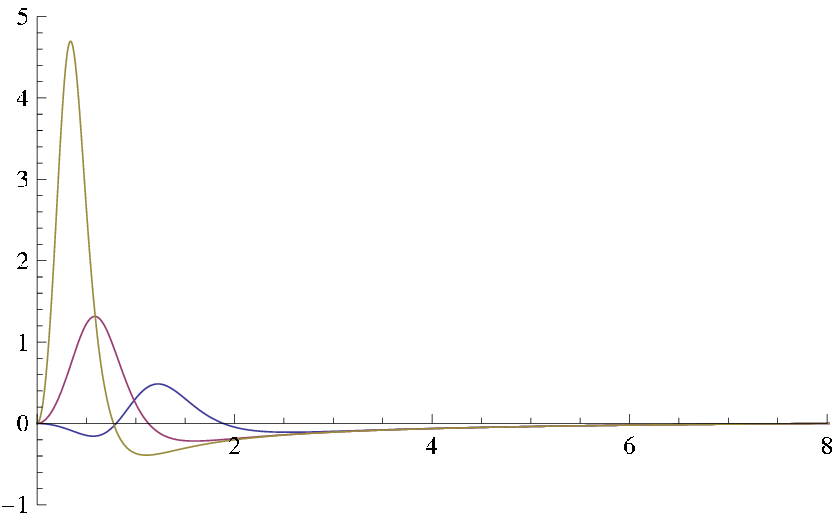}
\end{center}
\caption{The effective potentials $V(w)$ for the insulator models (\ref{Inss}),
(\ref{Soft}) and (\ref{Hard}) each corresponds to in the left, middle and right graph,
respectively. Each colored curve represents
different charge densities: $\rho=0.3$
(blue), $\rho=1$ (red) and $\rho=3$ (yellow).}
\label{potentialINSNEW}
\end{figure}

\begin{figure}[htbp]
\begin{center}
    \includegraphics[width=5cm]{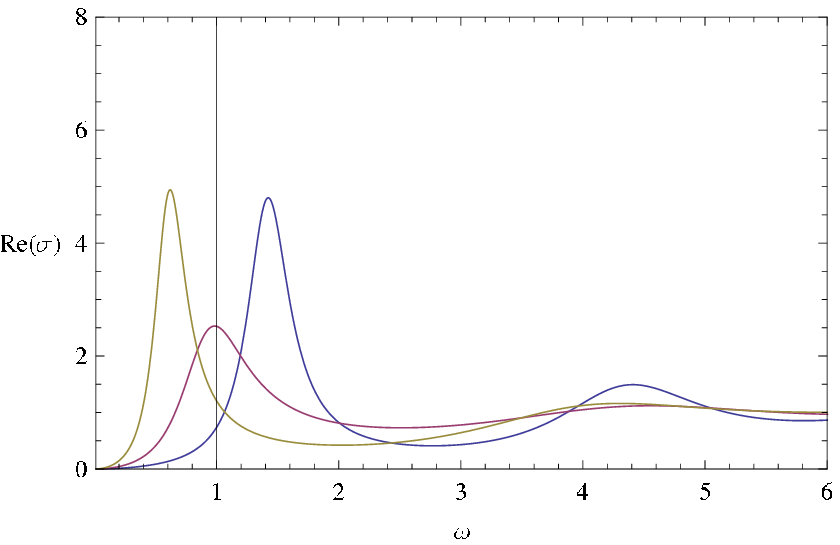}
\hspace*{0.5cm}
    \includegraphics[width=5cm]{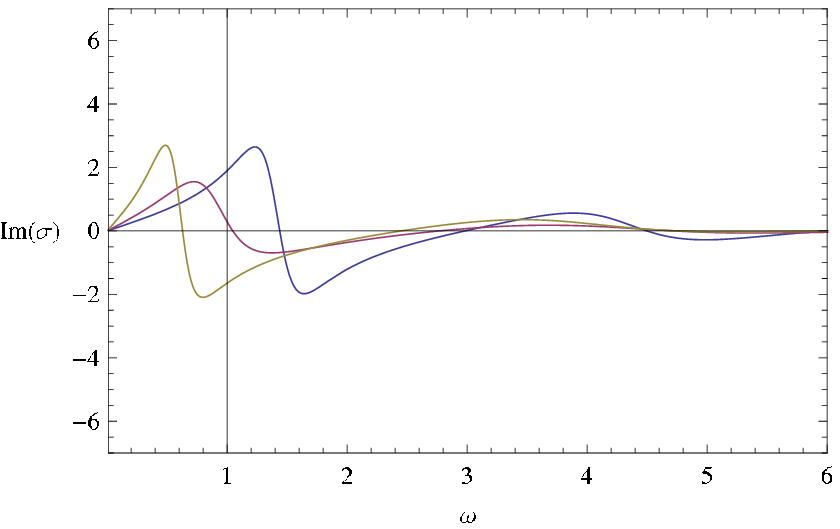}
\end{center}
\caption{The plot of real and imaginary part of the conductivity for the insulator model (i)
at $\rho=0.3$
(blue), $\rho=1$ (red) and $\rho=3$ (yellow).}
\label{condINSNEW}
\end{figure}

\begin{figure}[htbp]
\begin{center}
    \includegraphics[width=4.5cm]{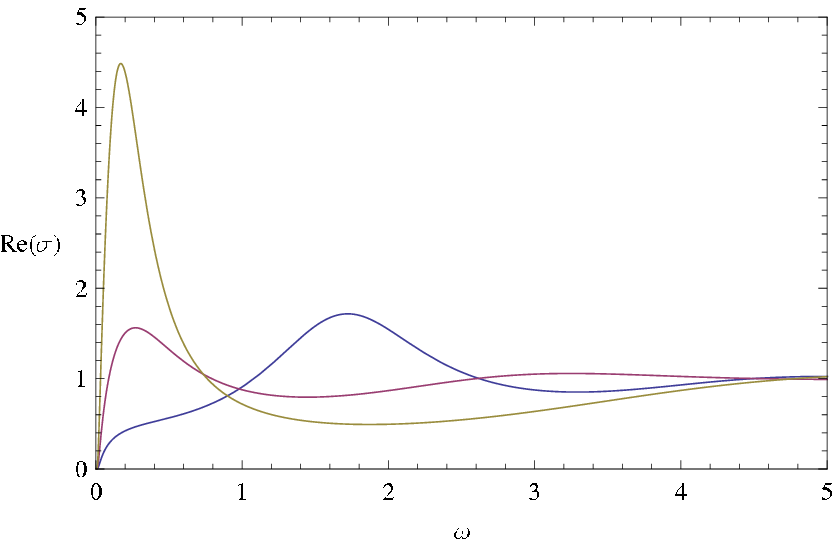}
    \includegraphics[width=4.5cm]{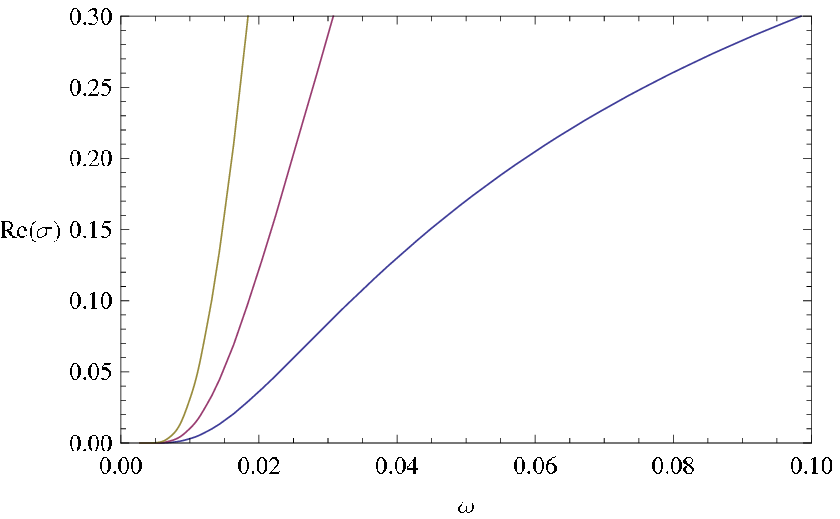}
\includegraphics[width=4.5cm]{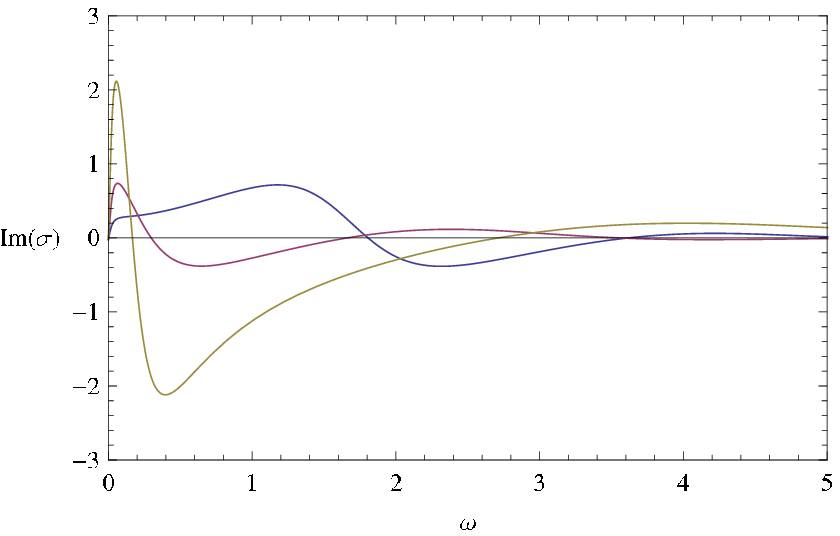}
\end{center}
\caption{The plot of real (the first and second graph) and imaginary part (the last one) of the conductivity for the insulator model (ii)
at $\rho=0.3$
(blue), $\rho=1$ (red) and $\rho=3$ (yellow).}
\label{condSINSNEW}
\end{figure}

\begin{figure}[htbp]
\begin{center}
    \includegraphics[width=5cm]{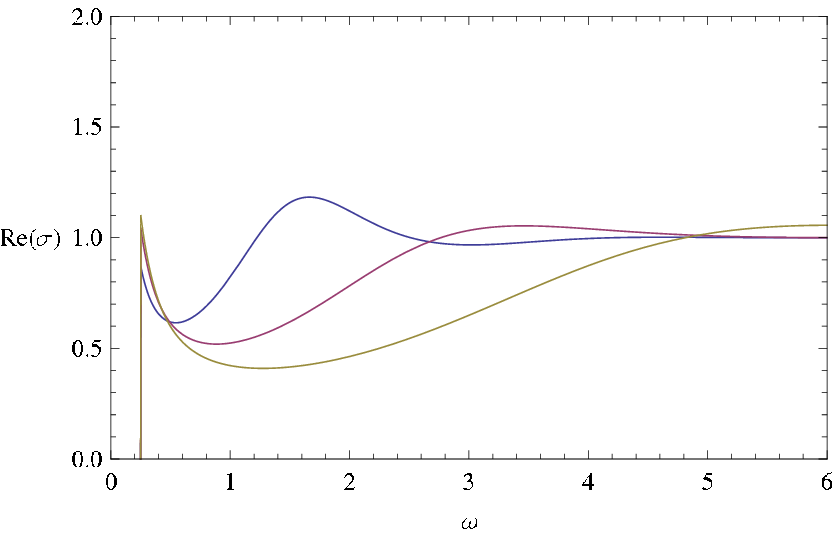}
\hspace*{0.5cm}
    \includegraphics[width=5cm]{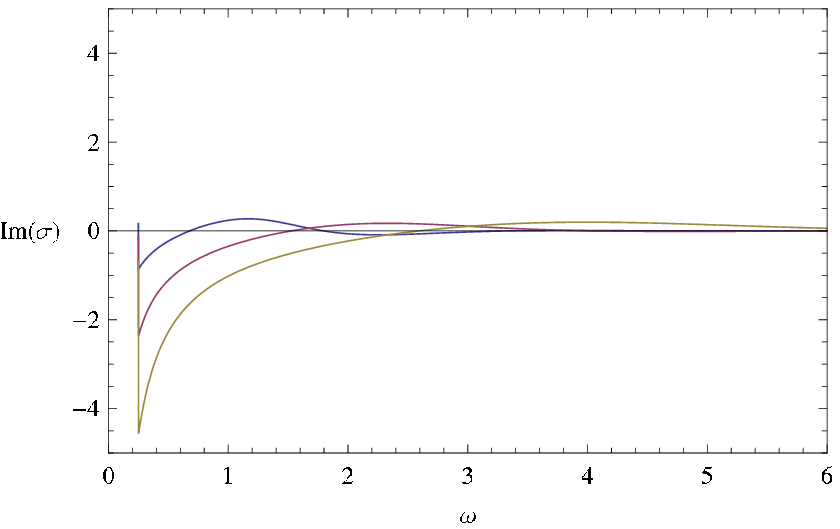}
\end{center}
\caption{The plot of real and imaginary part of the conductivity for the insulator model (iii)
at $\rho=0.3$
(blue), $\rho=1$ (red) and $\rho=3$ (yellow).}
\label{condHINSNEW}
\end{figure}

We will focus on the zero temperature limit, setting $h(z)=f(z)=1$ and analyze the following three different profiles $(i)-(iii)$ of the dilaton
\ba
&& (i)~ \mbox{Pseudo-gap insulator}:  e^{\phi(z)}=e^{\vp(z)}=1+z^8,   \label{Inss} \\
&& (ii)~ \mbox{Soft-gap insulator}: e^{\phi(z)}=e^{\vp(z)}=e^{(1+z^4)^\f{1}{8}}, \label{Soft} \\
&& (iii)~  \mbox{Hard-gap insulator}:e^{\phi(z)}=e^{\vp(z)}=e^{(1+z^4)^\f{1}{4}}. \label{Hard}
\ea
We leave explicit constructions of corresponding dilatonic backgrounds as supergravity solutions for a future problem. The effective potential for each background behaves as follows:
\ba
&& (i)~ \ \ V(w) \sim  O(w^2) \ \ (w\to 0),\ \ \ \ \ V(w)\sim   \f{2}{w^2}\ \ (w\to\infty), \no
&& (ii)~ \ \ V(w)\sim  O(w^2) \ \ (w\to 0),\ \ \ \ \ V(w)\sim   \f{1}{16w}\ \ (w\to\infty), \no
&& (iii)~ \ \ V(w)\sim  O(w^2) \ \ (w\to 0),\ \ \ \ \ V(w)\sim   \f{1}{16}\ \ (w\to\infty). \nonumber
\ea
The explicit forms of these potentials are plotted in Fig.\ref{potentialINSNEW}. The
holographic conductivity for each insulators are presented in Fig.\ref{condINSNEW}, Fig.\ref{condSINSNEW} and
Fig.\ref{condHINSNEW}, respectively.

In summary, we obtain the following behaviors of the AC conductivity at low frequency:
\ba
&& (i)~ \mbox{Pseudo-gap insulator}:\ \ \  \mbox{Re}\, \sigma(\omega)\to \omega^2 \ \ \ (\omega\to 0),   \label{cInss} \\
&& (ii)~ \mbox{Soft-gap insulator}: \ \ \  \mbox{Re}\, \sigma(\omega)\to e^{-\f{0.048}{\omega}} \ \ \ (\omega\to 0), \label{cSoft} \\
&& (iii)~  \mbox{Hard-gap insulator}:\ \ \  \mbox{Re}\, \sigma(\omega)=0 \ \ \ (0\leq \omega<\f{1}{4}). \label{cHard}
\ea
Notice that in all of these examples there are no delta functional Drude peaks and therefore they correspond to various kinds of insulators.

\clearpage

\section{Holographic Conductivity in Disordered Systems}\label{section:disorder}

So far we have analyzed the holographic conductivity in various probe D-brane setups.
In some sense, they are analogous to systems with electrons interacting with phonons and
owing to it
the DC conductivity gets finite. However, in realistic
materials, they usually include static impurities and they cause
the reduction of conductivity. The main purpose of this paper to incorporate the effect of impurities in the holographic frame work and this section will be devoted to this problem. We treat impurities as quenched disorder, which means time-independent randomness, as is standard in condensed matter physics \cite{Anderson,four} (see also e.g. reviews \cite{Disorder,DisorderInteractionReview,AndersonT}).

\subsection{Origins of Randomness in Probe Brane Setup}

We can consider several origins of randomness in our probe D-brane setups.
They are largely divided into two classes. One of them is the random fluctuations of the
background closed string fields (or supergravity fields)
such as the dilaton
$\phi(x,y,z)$ or the metric
$g_{\mu\nu}(x,y,z)$.
For example, if we regard the impurities as D-branes or other solitonic
objects in string theory localized in $x$ and $y$ direction and distributed randomly (similar to
the idea in \cite{KKY}), then
its backreaction clearly induces such random fluctuations of closed string fields.
This class is regarded as providing charge neutral impurities.
The other is those arising from
the background
open string fields, in particular the gauge potential $A_{t}$.
This is regarded as a charged impurity potential.
Also if the closed string
background fluctuates, then the solution (\ref{gpa}) with non-zero charge density
itself will be modified such that $A_t$ takes random values.
Since we are interested in how disorder affects low energy physics,
we can assume that the randomness disappears near the boundary $z=0$.
We will analyze these various disordered effects separately.
In Sections \ref{sec:dilaton} and
\ref{sec:charge}, we will study random fluctuations of charge density and
dilaton field, respectively.

An important issue when we consider the metric fluctuations is that
the horizon, which exists uniformly in a finite temperature gravity
dual, can locally disappear depending on the position of $(x,y)$ in
a random way. This means that the spacetime can locally change from
a AdS black brane to an AdS soliton. An AdS soliton denotes here an
asymptotically AdS spaces whose IR regions are capped off. A famous
example is the one which can be obtained from a double Wick rotation
of AdS black brane and this has been important in the context of
confinement/deconfinement phase transition \cite{WittenTh,HoMy}. The AdS
soliton itself behaves like a holographic insulator
\cite{Nishioka:2009zj}. This fluctuation, which makes a `hole' on
the horizon, turns out to be very important to understand
a transition from a metal into an insulator in our holographic setup as
we will see in section \ref{sec:hole}. This is because such a
hole is interpreted as an impurity which leads to a repulsive
interaction due to the mass gap dual to the AdS soliton geometry.
This effect can be approximately incorporated as the change of
boundary condition at the horizon as we will explain later.

Below we will analyze how the calculation of conductivity is changed due to the randomness.
 In order to obtain analytically exact results, we will first assume that the random background only depends on the coordinate $x$
but not on $y$, which corresponds to codimension one (or string shape) impurities.
 Finally, we will discuss the holographic conductivity when the randomness depends both on $x$ and $y$ in section \ref{sec:codim}.

\subsection{Definition of Conductivity in Disordered Systems}

  Before we proceed to the holographic calculation, it is important to make clear the definition of conductivity in disordered
systems. Below we will follow the argument in \cite{ReadC}. The presence of impurities breaks the translational invariance and thus
we encounter a bilocal conductivity $\sigma_{ij}(x,x',\omega)$ defined
by the linear response
between the current and electric field
\be
j_i(x,\omega)=\int dx' \sigma_{ij}(x,x',\omega) E_j(x',\omega),
\ee
where $i,j$ run over the space coordinates of a given system. Since in this paper we are only interested in the case $i=j=x$ and so we simply write $\sigma_{xx}$ as $\sigma$.

If we perform the random average denoted by $\la...\lb$, the translational invariance is recovered and we can write it as follows:
\be
\la \sigma(x,x',\omega)  \lb=\sigma(x-x',\omega).
\ee
Then we can define the AC conductivity in the disordered system as follows\footnote{
In this paper we employ the normalization of Fourier transformation such that
$A(x)=\f{1}{2\pi}\int dk e^{ikx}A(k)$ and $A(k)=\int dx e^{-ikx}A(x)$ for any function $A$.}
\be
\sigma(\omega)\equiv \sigma(k=0,\omega)=\int dx~ \sigma(x,\omega).
\ee

Now we would like to consider a simpler but equivalent definition of the conductivity in the disordered systems. We fix the external electric field as $E_x(x',\omega)=E_x(\omega)e^{ikx'}$ so that it has a
definite wave number $k$. In this case $j_x$ includes various wave numbers in general. However, after the disorder average we will get
\be
\la j_x(x,\omega)\lb=\sigma(k,\omega)E_x(\omega)e^{ikx}.
\ee
Thus we obtain for the wave number $k$ electric field
\be
\f{\la j_x(x,\omega)\lb}{E_x(x,\omega)}=\sigma(k,\omega).
\ee
By taking $k\to 0$,  we eventually find
\be
\sigma(\omega)=\f{\la j_x(x,\omega)\lb }{E_x(\omega)}, \label{dedconddis}
\ee
where notice that the random average $\la j_x(x,\omega)\lb$ actually does not depend on $x$.

\subsection{Analysis of Equations of Motion}

First we derive the equation of motion for the gauge field in DBI action for the general probe D-brane. In particular, we consider a generalized black brane background where the scalar fields $\phi$ and $\vp$ depend on both $x$ and $z$ and the metric is given by (\ref{metricp}) with $h(z)=f(z)$. We define a function $F(z,x)$ by generalizing the previous one (\ref{defp}) by (below we suppress the $R$ dependence by setting $R=1$ just for simplicity)
\be
\s{F(z,x)}\equiv\f{e^{-\vp(z,x)}}{\s{1-e^{-\phi(z,x)}z^4F_{tz}^2-\f{z^4e^{-\phi(z,x)}}{f(z)}F_{tx}^2
+z^4f(z) e^{-\phi(z,x)}F_{zx}^2}}.\label{deff}
\ee

The equations of motion of gauge fields from the DBI action (\ref{dbis}) are given as follows
for the $t$, $x$ and $z$ component:
\ba
&& \de_z(\s{F(z,x)}\de_z A_t)+\de_x\left(f(z)^{-1}\s{F(z,x)}(\de_xA_t-\de_tA_x)\right)=Q(x,z),\no
&& \de_t\left(f(z)^{-1}\s{F(z,x)}(\de_t A_x-\de_x A_t)\right)-\de_z(f(z)\s{F(z,x)} \de_z A_x)=0, \no
&& \de_t(\s{F(z,x)}\de_z A_t)-\de_x (f(z)\s{F(z,x)} \de_z A_x)=0,
\ea
where $Q(x,z)$ is the random charged source mentioned in the previous subsection.

It is clear that the static solution with a charge can be constructed in the form
(remember also the gauge condition $A_z=0$)
\be
A_t=\ap(x,z),\ \ A_x=0. \label{eleb}
\ee
The profile of $\ap(x,z)$ is found by solving
\be
\de_z(\s{F}\de_z \ap)+\de_x\left(f^{-1}\s{F}\de_x\ap\right)=Q(x,z), \label{indc}
\ee
with the standard boundary condition at the horizon
$\ap(z_H)=0.$ As we mentioned, it is sensible to assume that there is no randomness
at the boundary $z=0$ and thus we can always set
\be
F(0,x)=1.   \label{fcons}
\ee

Next we consider perturbations $A_t$ and $A_x$ around this solution to calculate the
conductivity.
The detailed form of the function $\ap$ is not important in the following
arguments. After we perform the Fourier transformation of $t$ into $\omega$,
the perturbative equations of motion read
\ba
&& f\de_z(f\s{F} \de_z A_x)+\left(\s{F}+\f{K}{f}(\de_x\ap)^2\right)
(\omega^2A_x-i\omega\de_xA_t)-i\omega K(\de_x\ap)(\de_z\ap)\de_zA_t=0, \no
\label{pemb}\\
&& \de_x (f\s{F} \de_z A_x)+i\omega\left(\s{F}+K(\de_z\ap)^2\right)\de_z A_t
+\f{K}{f}(\de_x\ap)(\de_z\ap)(-\omega^2A_x+i\omega\de_x A_t)=0. \no
\label{pemc}
\ea
We defined the function $K(z,x)$ by
\be
K(z,x)=\f{z^4e^{-\vp-\phi}}{\left(1-e^{-\phi}z^4F_{tz}^2-\f{z^4e^{-\phi}}{f}F_{tx}^2
+z^4 f e^{-\phi}F_{zx}^2\right)^{3/2}}. \label{kfun}
\ee
$F(z,x)$ and $K(z,x)$ in (\ref{pemb}) and (\ref{pemc}) are evaluated in the gauge field background (\ref{eleb}). Note that
they come from $x$ and $z$ component of the equations of motion of gauge fields and
we did not write explicitly that from $t$ component as it is not independent
from (\ref{pemb}) and (\ref{pemc}).

As we notice by plugging (\ref{eleb}) in (\ref{deff}) or (\ref{kfun}), in order to have a smooth
background, the function $\ap(z,x)$ should behave near the horizon as
\be
\ap(z,x)=\beta(x)(z_H-z)+O\left((z_H-z)^2\right) \ \ \ \ \ (z\to z_H), \label{apbc}
\ee
where the function $\beta(x)$ is a random function, which is determined by an explicit
random background we choose.

\subsection{Boundary Conditions}

Since the equations of motion are generally second order, we need
one boundary condition for each field. In black brane backgrounds, we
impose this at the horizon. For the gauge potential $A_t$ we require
\be A_t(z_H,x)=0, \label{atz} \ee as usual. Note that this
corresponds to a fixing of the residual gauge transformation $\delta
A_\mu=\de_\mu \chi(t,x)$. On the other hand, for the gauge field
$A_x$, we impose the in-going boundary condition at the horizon. By
using the condition (\ref{atz}) and equations (\ref{pemb}) and
(\ref{pemc}), we find that $A_x$ satisfies in the near horizon limit
$w\to\infty$
\be
\f{1}{\s{F}}\de_w (\s{F}\de_w A_x)
+\omega^2\left(1+\f{K(\de_x\ap)^2}
{f(\s{F}
+K(\de_z\ap)^2)}\right)A_x
+\f{K(\de_x\ap)(\de_z\ap)}{\s{F}(\s{F}+K(\de_z\ap)^2)}\de_x(\s{F}\de_w
A_x)=0. \label{eqaxh}
\ee
Moreover by using the condition
(\ref{apbc}), we can neglect terms proportional to $\de_x\ap$ in
(\ref{eqaxh}) in the near horizon. Therefore $A_x$ satisfied the
same equation as (\ref{axeqone}) and in-going boundary
condition is again given by (\ref{holbc}). By generalizing the
function (\ref{relax}), define \be X(w,x)=\f{\de_w
A_x(w,x)}{A_x(w,x)}, \label{relaxx} \ee and then we have the
integral expression \be A_x(w,x)=A_x(0,x)\cdot e^{\int^{w}_0 dw
X(w,x)}. \label{intrep} \ee Now the in-going boundary condition at
the horizon is simply written as \be X(\infty,x)=i\omega.
\label{bcxone} \ee

\subsection{Calculation of DC Conductivity}

After the previous preparations, we are now in a position to
holographically calculate the DC conductivity in quenched disordered
systems following the definition (\ref{dedconddis}). In the Fourier
basis, we normalize the external electric field $E_x=F_{tx}$ such
that it is given by $-i\omega$ at the boundary \be E_x(0,x)=-i\omega
A_x(0,x)-\de_xA_t(0,x)=-i\omega. \label{erel} \ee All we need to
solve are (\ref{pemb}) and (\ref{pemc}) under the boundary
conditions (\ref{atz}), (\ref{bcxone}) and (\ref{erel}). The AC
conductivity is obtained from \be \sigma(\omega)=T_p\left\la
\f{X(0,x)}{i\omega}\cdot A_x(0,x) \right\lb,  \label{conddd} \ee
where again $\la\ddd \lb$ denotes the random average.

For generic $\omega$, we have to resort to a rather complicated
numerical analysis to find $\sigma(\omega)$. However, if we are
interested in DC conductivity $\sigma(0)$ for codimension one
impurities, we can actually obtain an analytical result and this is
one of the main results in this paper.

Concentrating on the DC limit $\omega\to 0$, we can expand fields in a
series of $\omega$ as follows \ba && X(w,x)=i\omega\cdot
a(w,x)+O(\omega^2), \no && A_t(0,x)=i\omega\cdot
p(x)+O(\omega^2),\no && A_x(0,x)=1-\de_x p(x)+O(\omega),
\label{dclimit} \ea where we imposed (\ref{erel}). $a(w,x)$ and
$p(x)$ are functions which will be fixed by the equations of
motion.\footnote{Note that $p(x)$ cannot be eliminated by gauge
transformations because we impose the condition (\ref{atz}).}
The DC conductivity is now given, by taking the
limit $\omega\to 0$ of (\ref{conddd}),
by \be \sigma(0)=T_p\la
a(0,x)(1-\de_xp(x))\lb.  \label{dcconfd} \ee The boundary condition
(\ref{bdyconr}) is translated into \be a(\infty,x)=1.
\label{bdyconrr} \ee

In the DC limit (\ref{dclimit}), the equations of motion
(\ref{pemb}) and (\ref{pemc}) are simply reduced to \be
\de_w(\s{F}\de_w A_x)=\de_x(\s{F}\de_w A_x)=0. \ee Employing
(\ref{relaxx}) and (\ref{intrep}) in the DC limit, we find that the
combination \be \sigma_{DC}\equiv T_p\s{F(z,x)}\cdot
(1-\de_xp(x))\cdot a(w,x), \label{dcconst} \ee is a constant
\footnote{
One may wonder why the current $j_x$ and conductivity do
not depend on $x$ in spite of the $x$ dependent randomness. This is explained by the
current conservation $\f{\de j_x}{\de x }=-\f{\de\rho}{\de t}=O(\omega^2)$ if we
remember $\rho\sim \de_zA_t\sim O(\omega)$. This simple property is only true for
our codimension one impurities.} which
does not depend on $x$ and $z$.
Indeed, this quantity $\sigma_{DC}$
coincides with the DC conductivity we want to compute as is clear
from (\ref{fcons}) and (\ref{dcconfd}). By setting $z=z_H$ with the
condition (\ref{bdyconrr}), we have the expression \be
\sigma_{DC}=T_p\s{F(z_H,x)}\cdot (1-\de_xp(x)). \ee

To fix the value of the constant $\sigma_{DC}$, we need to look at the
behavior of the charge density $\rho$. As is clear from the
definition (\ref{chd}), it is perturbed by the linear fluctuations
which we exert in order to calculate the linear response. Thus this
perturbation of the charge density depends on $p(x)$. Therefore,
this physical consideration leads to the constraint require \be
\lim_{L\to\infty} \f{1}{L}\int_{0}^{L} \de_xp(x)=0,
\label{constc}\ee where $L$ is the length in the $x$ direction and
should be taken to be infinity. This relation  says that the charge
density should not change a lot in the presence of the disorder and
is clear if we impose a periodic boundary condition $p(x+L)=p(x)$ or
a damping condition at the boundaries $x=0$ and $x=L$. This requirement (\ref{constc})
allows us to obtain the final expression of the DC conductivity
after taking the random average: \be
\sigma_{DC}=T_p\left\la\f{1}{\f{1}{L}\int^L_0 dx
\f{1}{\s{F(z_H,x)}}}\right\lb\ . \label{finalcon}
 \ee
Notice also that the formula (\ref{finalcon}) is reduced to the previous result (\ref{condx})
if we turn off the randomness.

Since the function $\f{1}{x}$ is convex, we can easily see \be
\sigma_{DC}\leq T_p\left\la \f{1}{L}\int^L_0 dx
\s{F(z_H,x)}\right\lb.   \label{conv} \ee Notice that the right-hand side of the
above inequality is a naive average of conductivity based on the
result (\ref{condx}) without disorder. From this general argument it is already clear that the presence of
impurities tends to suppress the conductivity in our holographic calculations.

\subsection{Random Dilaton: Charge Neutral Impurities} \label{sec:dilaton}

Now we would like to study the DC conductivity (\ref{finalcon}) in more detail.
First we would like to understand the effect of random values of the dilaton field $\phi$.
As is clear from (\ref{condx}), a large value of the dilaton reduces the conductivity.
For a large charge density $\rho z^2_H/R^2 >>1$, we find
\ba
\sigma_{DC}\simeq T_p\rho z^2_H\left\la\f{1}{\f{1}{L}\int^L_0 dx~
e^{\phi(z_H,x)}}\right\lb\ . \label{disdilaton}
\ea

In particular, if the region where the dilaton gets divergent
has a small but non-zero volume, then the DC conductivity vanishes
$\sigma_{DC}=0$ as is clear from (\ref{disdilaton}). This holographically
argues that the originally metallic system can be changed into an insulator
in the presence of impurities. Notice that the
right hand side of (\ref{conv}), which corresponds to
a naive average, remains non-zero in this case.

In this way, we find
that the dilatonic disorder effect highly reduces the DC conductivity as we
expect from the standard knowledge for weakly coupled condensed
matter systems. Notice that our results are obtained using
holography with a supergravity approximation and thus the dual gauge
theories are strongly coupled, which is usually very difficult to
analyze in condensed matter physics.

One may worry that in the above argument,
the divergent value of dilaton is important to realize insulators. However, we will
later show in section \ref{sec:hole} that it is still possible to obtain insulators
in a different way without assuming singular backgrounds.

\subsection{Random Charge Density: Charged Impurities} \label{sec:charge}

Next we study the effect of the random charge density $\rho(x)$, which induces random fluctuations of
gauge potential in (\ref{eleb}). To extract only this effect, we turn off the
scalar fields $\phi=\vp=0$ below.
We start with a perturbative argument assuming that the randomness is small. Consider a fluctuation
$\delta \beta(x)$ of the function $\beta(x)$ in (\ref{apbc}) from the background charged solution
(we revived $R$ dependence)

\be
\beta(x)=-\f{\rho_0}{\s{1+\f{\rho_0^2 z^4_H}{R^4}}}+\delta\beta(x),
\ee
where $\rho_0$ is the original charge density.
By evaluating the expression (\ref{deff}) perturbatively
\be
F(z_H,x)=\f{1}{1-\f{z^4_H\beta^2(x)}{R^4}}>1,\label{exprf}
\ee
in the end we find the leading correction due to the disorder effect
\ba
\Delta\sigma_{DC} &=& \sigma_{DC}-T_p\s{1+\rho_0^2z_H^4}\no
&=& \f{T_p z^4_H}{2R^4}\left(\f{1}{L}\int^L_0 dx \left(1+\f{\rho_0^2z^4_H}{R^4}\right)^{5/2}\la(\delta\beta(x))^2\lb\right)\no
& & +\f{T_p\rho_0^2 z^8_H}{R^8}\left(1+\f{\rho_0^2z^4_H}{R^4}\right)^{3/2}
\left\la\left(\f{1}{L}\int^L_0 dx \delta\beta(x)\right)^2\right\lb\geq 0.
\label{antil}
\ea

\begin{figure}[htbp]
\begin{center}
    \includegraphics[width=6cm]{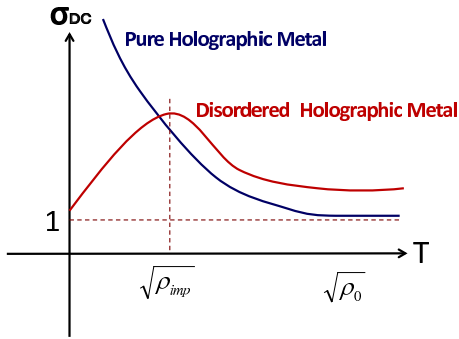}
\end{center}
\caption{The schematic plot of the DC conductivity as a function of temperature in pure and disordered
holographic metals with a random charge density. We simply set $T_p=R=1$.}
\label{hdiscondfig}
\end{figure}

In this way,
we find that the effect of the disordered charge
density profile increases the conductivity when its
random fluctuation is small. However, as we will explain this slightly non-standard behavior changes if the randomness increases or the temperature decreases.
To see this, we come back to the expression (\ref{exprf}). We can regard the amplitude of fluctuations of $\delta \beta(x)$ as the impurity density $\rho_{imp}(<<\rho_0)$. The previous perturbation
breaks down when the temperature gets smaller than (remember the relation $T\sim\f{1}{z_H}$)
\be
T_c\sim \f{\s{\rho_{imp}}}{R}.  \label{tempi}
\ee
Below this temperature, $\beta(x)$ is dominated by the random perturbation
$\delta\beta(x)$ and therefore $\sigma_{DC}$ will decrease from the value (\ref{condx}) $\f{\sigma_{DC}}{T_p}\sim \f{\rho}{T_c^2R^2}\sim \f{\rho}{\rho_{imp}}>>1$ to $\f{\sigma_{DC}}{T_p}\simeq 1$. Notice that $\f{\sigma_{DC}}{T_p}$ is always greater
than $1$ as is clear from (\ref{exprf}) and (\ref{finalcon}) because here we did not turn on the random fluctuations of closed string fields. These behavior is sketched in
Fig.\ \ref{hdiscondfig}.

This somewhat reminds us of what happens
typically for the DC conductivity
in systems with relativistic dispersion
(e.g., graphene, quasi-particles in the two-dimensional $d$-wave superconductor,etc)
where the disorder effect on the quantum transport can be interpreted in a twofold way.
When the chemical potential is close to the Dirac point,
disorder can increase the density of states (DOS),
leading to enhancement of DC conductivity,
which is however in competition with
the diffusion constant diminished by disorder.
As a consequence of the delicate balance between these,
when the chemical potential is exactly
at the Dirac point,
a leading order effect of disorder potential that we see
in, say, the self-consistent Born approximation or semiclassical analysis,
often gives rise to the DC conductivity which is not affected by disorder.
(However, as we increase disorder or as we lower temperature,
we need to include higher order effects or quantum effects from disorder potential.
In particular, these higher-order or non-perturbative effects
in graphene include
the Berry phase and topology associated to the single particle wavefunctions in graphene.
Because of the topological reason,
when the charge impurity potential does not change at atomic scales,
disorder never be able to localize electrons
but instead it {\it increases} the DC conductivity
\cite{Ostrovsky07, Ryu2007, Bardarson07, Nomura2007}.
)
With interactions and with phonon bath
which are included, together with disorder, in our holographic
calculation in non-perturbative way,
these competing effects are not in proportion any more
and enhancing the conductivity.

In this setup we can have two mean free times
(or relaxation times), each for the elastic scattering
$\tau_{el}$ and for the inelastic scattering $\tau_{in}$. The former is due to the presence of impurities
 and we can estimate $\tau_{el}\sim \f{1}{\s{\rho_{imp}}}$.
On the other hand,
 the latter is due to the interactions mediated by gluons and is related to the temperature as
 $\tau_{in}\sim 1/T$.
Thus, as we decrease temperature, $\tau_{in}$ exceeds $\tau_{el}$
around the critical temperature $T_c$ (\ref{tempi}).

\subsection{Random Holes on Horizon} \label{sec:hole}

\begin{figure}[htbp]
\begin{center}
    \includegraphics[width=5cm]{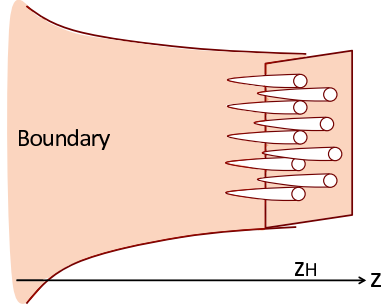}
\end{center}
\caption{A sketch of holes on a black brane horizon which are dual to impurities in holographic metals.}
\label{holefig}
\end{figure}

In the previous formulation, we took into account the random
fluctuations of the scalar fields $\phi$ and $\vp$ and the gauge
potential $A_t$. One may think that we should also include those of
the metric $g_{\mu\nu}$. In this paper we will not do this literally
as calculations get highly complicated. Instead, we approximately
incorporate the effect of large perturbations which make small but
many 'holes' on the horizon\footnote{Numerical solutions of
localized black holes inside a AdS soliton have been constructed in
\cite{AMW}.} by locally replacing the black brane with a solitonic
geometry, where the IR region is capped off (see Fig.\ref{holefig}).
As we mentioned, this introduces
a mass gap locally and each hole is
expected to describe holographically an impurity potential.

The situation can be compared with relativistic fermions subjected to
``random mass'' type disorder.
It is described by the following single particle Hamiltonian:
\begin{eqnarray}
\mathcal{H}
\!\!&=&\!\!
-{i} \boldsymbol{\sigma}\cdot  \boldsymbol{\partial}
+
m(x,y) \sigma_z,
\end{eqnarray}
where $\boldsymbol{\partial}=(\partial_x,\partial_y)$,
and $\sigma_{x,y,z}$ denotes the $2\times 2$ Pauli matrix;
$m(x,y)$ represents a random mass term.
Such a random mass Dirac problem was discussed
in the context of e.g.,
the quantum Hall plateau transition, the random bond Ising model,
and graphene (see, for example, \cite{Ludwig94}, \cite{Cho97},
\cite{Nomura2008}
etc. and references therein).
We should keep in mind, however, that our holographic calculation
includes effects of interactions and heat bath on top of the disorder potential.

Instead of dealing with explicit metric, we will describe
this situation qualitatively by changing the in-going boundary condition locally as
\ba
A_x(w,x)\to a(x)\left(e^{i\omega (w-w_{\infty})}+R(x)e^{-i\omega (w-w_{\infty})}\right), \ \ \
(w\to w_\infty) \label{bdym} \ea
where $w_\infty$ is the IR cut off at which the tip of the AdS soliton located
and $a(x)$ is an arbitrary function.
The function
$R(x)$ interpolates between $R(x)=0$ and $R(x)=1$,
where $R(x)=0$ corresponds
to the regular point where the impurity potential is negligible (or equally the in-going
boundary condition),
and $R(x)=1$ corresponds to the center of
impurity (or equally the Dirichlet
boundary condition).

This randomly
modified boundary condition is simply expressed as
follows\footnote{Strictly speaking we need to
shift slightly the
position from the horizon to impose (\ref{bdyconr}). This is because
if we literally impose (\ref{bdyconr}), the order $\omega^2$ term of
(\ref{eqaxh}) gets divergent if $\theta(x)\neq 1$ when we integrate
$w$ toward $\infty$ to find a solution of $A_x$. However, this
divergence is artificial because we will not get the horizon
geometry in the case of AdS soliton and we need to introduce another
IR cut off $z_{imp}>z_H$ which is dual to the mass gap due to the
impurity. Thus we actually find that the $\omega^2$ term always
remains finite and can be negligible in the DC limit $\omega\to 0$.}
\be X(\infty,x)=i\omega\cdot \theta(x), \label{bdyconr} \ee
where we defined $\theta(x)=\f{1-R(x)}{1+R(x)}$. Points with $\theta(x)=0$
describe the impurities, while those with $\theta(x)=1$ do the metallic region.
 Notice that this is a description for low temperature
systems and as we increase the temperature, the region with
$\theta(x)=0$ will gradually disappear.

By employing this disordered boundary
condition (\ref{bdyconr}), the calculation of the conductivity can be done in the same way as
we did in the previous section. The final formula (\ref{finalcon})
is simply replaced with \be
\sigma_{DC}=T_p\left\la\f{1}{\f{1}{L}\int^L_0 dx
\f{1}{\theta(x)\s{F(z_H,x)}}}\right\lb\ . \label{finalcont}
 \ee

In this setup, we can realize the insulator $\sigma_{DC}=0$ by
keeping the dilaton finite and staying with the regular solutions of
supergravity. For example, if we assume the following profile \be
\theta(x)\s{F(z_H,x)}=\f{(x-L/2)^2+\ep^2a^2}{(x-L/2)^2+a^2}, \label{thpf}\ee then we find \be
\sigma_{DC}=\f{1}{1+\f{\pi a}{\ep L}(1-\ep^2)}, \ee assuming $L\gg
\ep$. Thus if we take the limit $\ep=0$, then $\sigma_{DC}$
vanishes. More generally, if the function $\theta(x)$ has  at least
one generic zero $x=x_0$ i.e. $\theta(x_0)=0$ with $\theta(x)\propto
(x-x_0)^2$ in the limit $x\to x_0$, then the DC conductivity
vanishes. This shows that even a single impurity can turn a metal
into an insulator in $1+1$ dimensions.

It is known that a similar thing happens in the Tomonaga-Luttinger liquid
when there are predominantly repulsive electron-electron interactions
\cite{KaneFisher91, Furusaki93}.
In this case,
the local potential is a relevant operator in the RG sense
and
at low energies (at low temperature),
the single impurity effectively cuts the systems into two
separate parts.

Before we go on, we would like to remember again that to obtain the result (\ref{finalcont}),
we assumed codimension one impurities. Therefore, more naturally, this corresponds
to point-like impurities in $1+1$ dimensional system described by
AdS$_3$/CFT$_2$. In this case the DC conductivity is given by a similar formula
\be
\sigma_{DC}=T_pz_H\left\la\f{1}{\f{1}{L}\int^L_0 dx
\f{1}{\theta(x)\s{F(z_H,x)}}}\right\lb\ . \label{finalconthr}
 \ee

\subsection{Codimension Two Randomness} \label{sec:codim}

So far our holographic calculations are restricted to the codimension one impurities by
assuming that the randomness only depends on $x$ but not on $y$. Even though this is enough
to describe $1+1$ dimensional disordered systems, for realistic setups of $2+1$ dimensional systems,
usually the impurities are point-like (i.e. codimension two) and thus we need to further
 incorporate the $y$ dependence.

If we allow both $F$ and $\theta$ to depend on both $x$ and $y$, the equation of motion gets highly complicated. Thus here we assume that $F=F(z)$ is a constant function of $x$ and $y$, while $\theta=\theta(x,y)$ is a random function. This means that we ignore fluctuations of the gauge field. Thus our result below can give full contributions only when the charge density is vanishing $\rho=0$.
In this case, the equations of motion of the gauge fields take the following form (each comes from $x$, $y$ and $z$ component) to the leading order
\ba
&& \de_w(\s{F}\de_w A_x)=\de_y(\ddd), \label{neomx} \\
&& \de_w(\s{F}\de_w A_y)=\de_x(\ddd),\label{neomy} \\
&& \de_x(\s{F}\de_w A_x)+\de_y(\s{F}\de_w A_y)=0, \label{neomz}
\ea
 where $(\ddd)$ represent certain functionals of the gauge field whose details are not
 important. Assuming an appropriate boundary condition in $y$ direction as before,
 we can see from (\ref{neomx}) and (\ref{neomz}) that
 \be
\sigma_{DC}=\f{T_p}{L} \int^L_0 dy \s{F(z)}~\f{\de_w A_x(z,x,y)}{i\omega}, \label{dcxy}
 \ee
does not depend on $x$ and $w$ and therefore coincides with the DC conductivity.

By generalizing (\ref{dclimit}), we impose the boundary behavior
\ba
&& A_t(0,x,y)=i\omega p(x,y) +O(\omega^2),\no
&& A_x(0,x,y)=1-\de_x p(x,y) +O(\omega^2),\no
&& A_y(0,x,y)=-\de_y p(x,y) +O(\omega^2),\no
&& X(z_H,x,y)=i\omega\theta(x,y).
\ea
Then, the equation (\ref{neomz}) evaluated at the horizon leads to
\be
\de_x\left[\theta(x,y)(1-\de_x p(x,y))\right]-\de_y[\theta(x,y)\de_yp(x,y)]=0. \label{eomxy}
\ee
Moreover, we need to impose the boundary condition such that the average of the
gradient of the charge density is vanishing as we did
before in (\ref{constc}). One choice will be to require
\be
p(x,y)=0, \label{bcxy}
\ee
on the boundary of the region with the linear size $L$.
After solving (\ref{eomxy}) with this boundary condition, the DC conductivity is finally obtained from (\ref{dcxy}) setting $z=z_H$:
\be
\sigma_{DC}=\f{T_p\s{F(z_H)}}{L}\int^L_0 dy~\theta(x,y)(1-\de_x p(x,y)). \label{dcfinf}
\ee
For simplicity, we suppress the coefficient $T_p\s{F(z_H)}$ below, by setting this to be one.

In general, we cannot expect analytical solutions to the present
problem (\ref{eomxy}) and (\ref{bcxy}). However, it is helpful to
rewrite (\ref{eomxy}) into a form like Schrodinger problem by
defining \be \vp(x,y)=\s{\theta(x,y)}p(x,y). \ee This satisfies the
following Schrodinger equation with a source: \be
-(\de^2_x+\de^2_y)\vp+V(x,y)\vp=W(x,y), \label{schrd} \ee where \ba
&&
V(x,y)=\f{(\de^2_x+\de^2_y)\theta}{2\theta}-\f{(\de_x\theta)^2+(\de_y\theta)^2}{4\theta^2},
\no && W(x,y)=-\f{\de_x\theta}{\s{\theta}}. \ea

\begin{figure}[htbp]
\begin{center}
    \includegraphics[width=5cm]{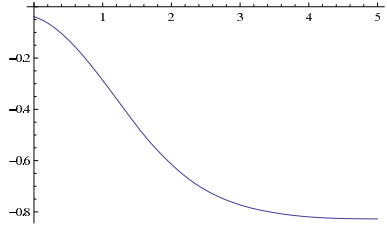}
\end{center}
\caption{The value of $L^2\cdot\Delta\sigma$ defined in (\ref{xydd}) versus the value of $\log L$.}
\label{listplot}
\end{figure}

As an example, we model a single impurity situated at $x=y=0$ by the
profile \be \theta(x,y)=\f{r^2}{r^2+1}, \ee where we employ the
polar coordinate $x=r\sin(s)$ and $y=r\cos(s)$. Also we obtain \be
V(r)=\f{1-2r^2}{r^2(1+r^2)^2},\ \ \ \
W(r,s)=-\f{2\sin(s)}{(1+r^2)^{3/2}}. \ee
We consider the system which is
given by a disk with radius $L$
and the boundary condition (\ref{bcxy})
leads to \be \vp(r=L)=0. \ee We can assume the following form of
solutions \be \vp(x,y)=\vp(r)\sin(s). \ee The equation (\ref{schrd})
is written explicitly as
\be
-\de^2_r\vp-\f{\de_r\vp}{r}+\left(V(r)+\f{1}{r^2}\right)\vp=-\f{2}{(1+r^2)^{3/2}}.
\ee This equation with the smoothness condition at $r=0$ requires
the behavior $\vp(r)\propto r^{\s{2}}$ and thus $p(x,y)$ behaves like
\be p(x,y)=(\mbox{const.})\cdot r^{\s{2}-1}\sin(s),\ \ \ (r\to 0).
\ee The conductivity is expressed as \be
\sigma_{DC}=\sigma_0+\Delta\sigma, \ee where \ba &&
\sigma_0=\f{2}{L^2}\int^L_0 dr~r\theta(r),\no &&
\Delta\sigma=-\f{1}{L^2}\int^L_0 dr~r\theta(r)
\left(\f{\vp(r)}{r\s{\theta(r)}}+\f{\de}{\de
r}\left(\f{\vp(r)}{\s{\theta(r)}}\right)\right).\label{xydd} \ea We
numerically confirmed that the difference $\Delta\sigma$ between the
true DC conductivity $\sigma_{DC}$ and its naive value $\sigma_0$ is
negative for any choice of $L$ as plotted in Fig.\ \ref{listplot} and
this property is similar to the previous codimension one case. As it
shows, it approaches to $\Delta\sigma\simeq -\f{0.831}{L^2}$ in the
$L\to\infty$ limit.

Notice that in this holographic calculation for the codimension two impurity, $\Delta\sigma$ cannot
be order one as opposed to the previous codimension one case. Therefore in $2+1$ dimension we can conclude that  a single or finite number of impurities cannot make a metal into an insulator, as expected in condensed matter physics.

Another simple example will be the case where
$\theta$ depends only\footnote{
Note that if $\theta$ depends only on $x$, the system is reduced to the system discussed in the previous subsection.}
 on $y$. Because the source term $W$ in (\ref{schrd}) vanishes, we can set $\vp=p=0$. Thus we find that the naive conductivity coincides with $\sigma_{DC}$
 \be
 \sigma_{DC}=\f{1}{L}\int^L_0 dy \theta(y).
  \ee
This shows that $\sigma_{DC}$ is non-vanishing unless $\theta(y)=0$
identically.

\begin{figure}[htbp]
\begin{center}
 \includegraphics[width=6cm]{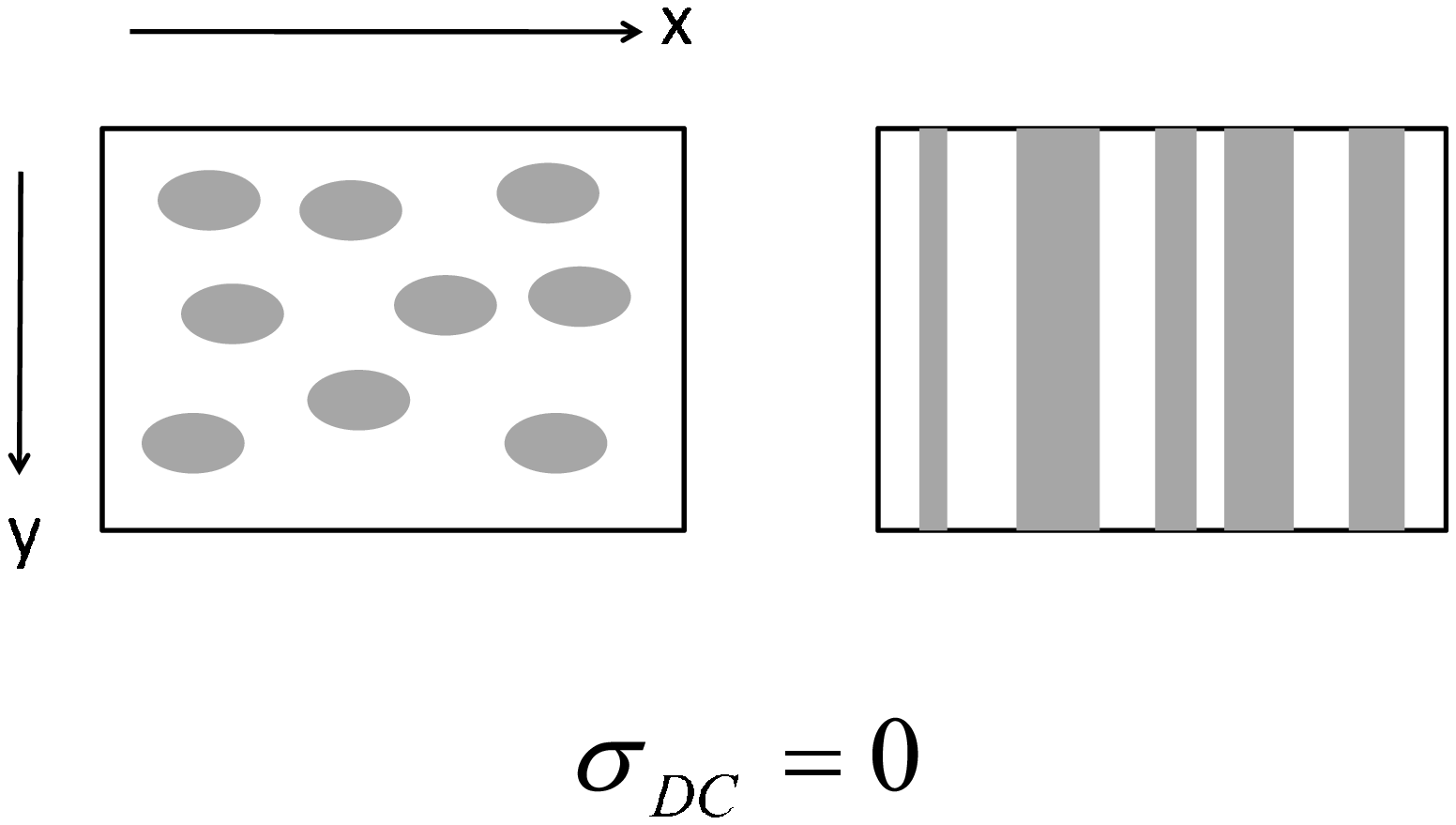}
 \hspace{2cm}
    \includegraphics[width=6cm]{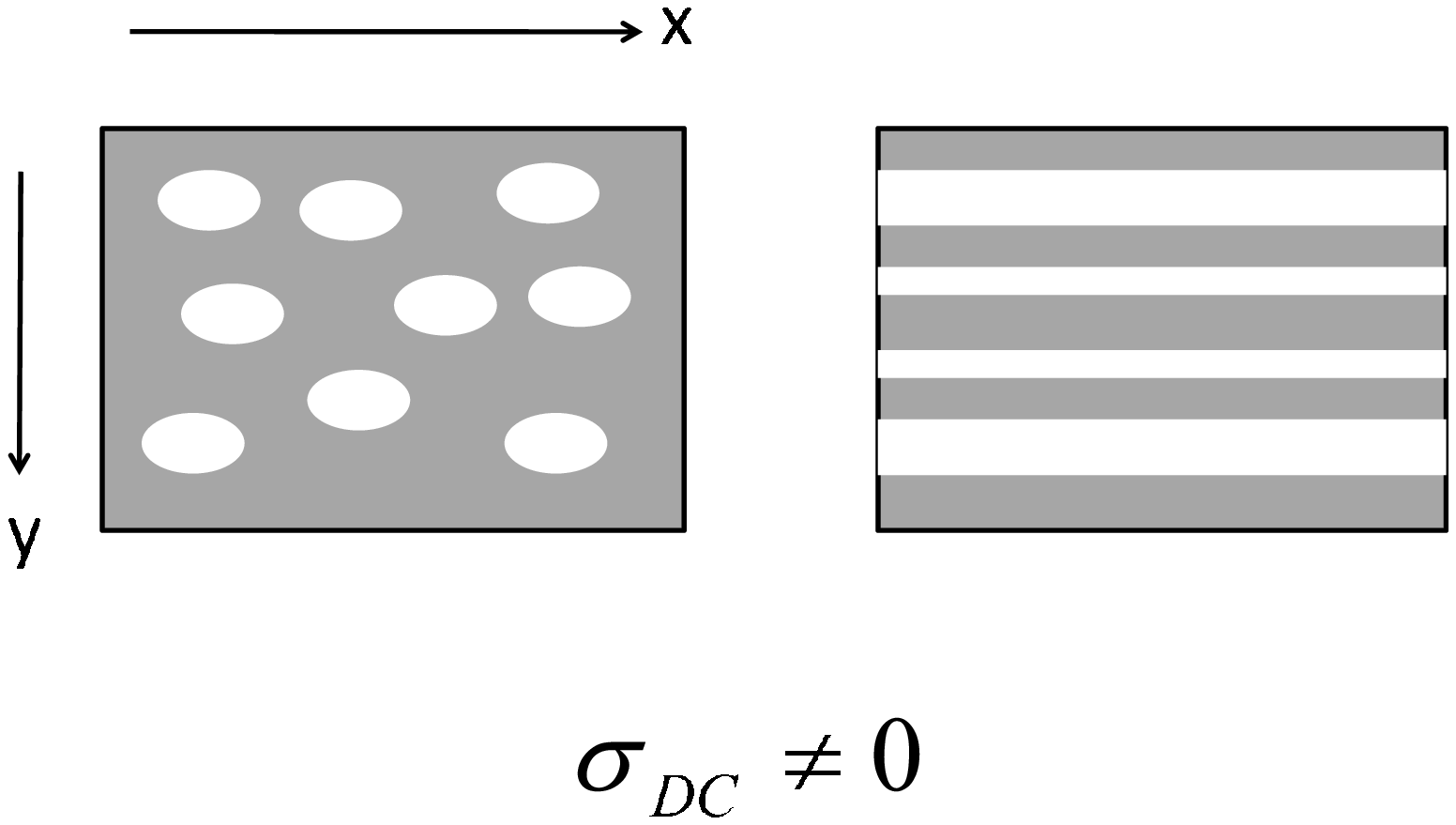}
\end{center}
\caption{Schematic examples of the holographic conductivity. The
white regions represent the ones where $\theta=0$. The gray regions
are the ones where $\theta>0$. The left two figures correspond to an
insulator, while the right ones to a metal.} \label{condtwo}
\end{figure}

One of the most interesting questions in this system will be whether
it shows a metal-insulator phase transition as we increases the
number of impurities. Even though we need a full numerical analysis
to answer this question completely, our previous analysis suggested
that the conductivity becomes zero only if there is no connected
path between $x=0$ and $x=L$ on which $\theta$ is non-vanishing. For
example, if $\theta$ vanishes at $x=x_0$ for any values of $y$, the
conductivity vanishes as we showed in the previous section. For
schematic examples, see the Fig.\ \ref{condtwo}.

Finally it is intriguing to notice that in the presence of random
distributions of $\theta(x,y)$, the potential $V$ becomes a random
function of $x$ and $y$ and our problem to solve (\ref{schrd}) looks
similar to the problem of Anderson localization, which is also
described by a random potential.

\section{Conclusions and Discussions}

In this paper we studied the holographic conductivity in various
probe brane systems via the AdS/CFT
correspondence. In the first half of this paper, we analyzed rather
general setups of probe D-branes which are dual to 2+1 dimensional systems.
They are flavor or defect
sectors of gauge theories, which are coupled to gluons in the
same or higher dimensions.
Our formalism incorporates the effect of dilaton which depends on
the radial coordinate of the AdS.
We present a systematic analysis of AC and DC conductivity by
relating this to the Schrodinger problem.
As one of the examples, we found that a known dilatonic solution,
which can be embedded in string theory, leads to a peculiar
pseudo-gap of the form
$\mathrm{Re}\,\sigma(\omega)\sim \omega^{1/3}$ in addition to
the delta functional Drude peak at zero temperature. In order
to realize insulators, we somehow need to remove the Drude peak.
Since the DC conductivity depends on the value of dilaton at the
horizon, we noted that a system with a divergent dilaton can be
regarded as a holographic insulator, which resembles Mott
insulators. We confirmed this interpretation by calculating AC
conductivity numerically and found that we can realize various
insulators by choosing the dilaton profile appropriately.

In the latter half, we concentrated on the main issue of this paper.
We studied the holographic conductivity in disordered systems.
In particular, we again employed the probe D-brane setups
because they allow us to calculate the effect of randomness in the
classical analysis and because we can ignore complicated backreactions.
We consider three important origins of the disorder. Two of them
are closed string fields: dilaton and the metric. We treat the
metric fluctuations as holes on the horizon, assuming a low temperature region.
The other is an open string field: the $U(1)$ gauge field on the probe
D-brane, which corresponds to the
random charged density. We showed that the dilaton and metric
randomness can reduce the DC conductivity and can finally take
it to be zero. On the other hand, the disordered charge density
increases the conductivity at high temperature and as the temperature
goes down, its effect changes the sign and eventually drives it to a
certain small but non-zero value.
We also managed to obtain an analytical formula of DC conductivity
 for the
codimension one randomness. In more general cases, we gave a perturbative
 estimation in the presence of the
disordered closed string fields.

Finally, one of the most intriguing questions will be whether our
holographic analysis dual to strongly coupled disordered systems
leads to a phenomenon similar to Anderson localization, which has
been usually derived for weakly coupled electrons. For example, let
us choose the characteristic values at finite temperature $T\sim
1/z_H$  in the profile  (\ref{thpf}) as $a\sim T^{-1}$ and $L\sim
T^{-1}$. Assuming that the disorder effect is very small at high
enough temperature, we can treat $\delta\equiv 1-\ep^2$ as a very
small parameter $\delta\ll 1$. In the AdS$_3$ case dual to one
dimensional systems, the DC conductivity (\ref{finalconthr}) is
perturbatively estimated to be
$\sigma_{DC}=\sigma^{pure}_{DC}-C\cdot\f{\delta}{T}+O(\delta^2)$,
where $\sigma^{pure}_{DC}$ is the conductivity without disorder and
$C$ is a positive numerical constant. This dependence on the
temperature seems to be consistent with the known formula in weak
localization \cite{four,Disorder} for one dimensional systems.
However, a similar analysis in AdS$_4$ case does not seem to
reproduce the known logarithmic dependence on the temperature.
Therefore we cannot argue that our holographic analysis in
supergravity approximation, which is dual to the strong coupling
limit and large N limit of CFT, support the Anderson localization.
 We hope we can come back to more details in future publications.

\vspace{1cm}

\noindent {\bf Acknowledgments} We are grateful to Hong Liu and
Elias Kiritsis for useful conversations. SR is supported by Center
for Condensed Matter Theory at University of California, Berkeley.
TT and TU are supported by World Premier International Research
Center Initiative (WPI Initiative), MEXT, Japan. SR and TT would
like to thank the winter conference ``Strongly Correlated Systems
and Gauge/Gravity Duality,'' held in 2011 at Aspen center for
physics, where a part of this work was completed. TT is grateful to
the workshop ``AdS/CM duality and other approaches,'' held in 2010
at KITPC, the Chinese Academy of Sciences, where an initial stage of
this work has been progressed. The work of TT is also supported in
part by JSPS Grant-in-Aid for Scientific Research No.20740132, and
by JSPS Grant-in-Aid for Creative Scientific Research No.\,19GS0219.


\appendix

\section{Sum Rule}\label{setion:sumrule}

Since the sum rule is useful to understand the general behavior of the conductivity, here we would like to summarize it. The general aspect of sum rules in AdS/CFT has been considered in (\cite{Gulotta:2010cu}). Here we would like to concentrate on that of conductivity in the $2+1$ dimensional critical system and study its consequence.

If the retarded Green function $G_R(\omega)$ is holomorphic for
$\mathrm{Im}\,\omega\geq 0$ and it satisfies
$\lim_{|\omega|\to 0}G_R(\omega)\to 0$ for $\mathrm{Im}\,\omega\geq 0$, then
\be
G_R(0)=\lim_{\ep\to 0}\int^\infty_{-\infty} \f{d\omega}{\pi}\f{\mbox{Im}G_R(\omega)}{\omega-i\ep}.
\ee
Since Im$G_R(\omega)$ is an odd function of $\omega$, we can replace the integral in (\ref{sumr}) with
\be
G_R(0)=2P\int^\infty_{0} \f{d\omega}{\pi}\f{\mbox{Im}G_R(\omega)}{\omega}.
\label{sumr}
\ee

Now we remember the relation in linear response theory
\be
\sigma(\omega)=\f{\la J_x(\omega) J_x(0)\lb_R}{\omega}.
\ee
In 2+1 dimensional systems which are critical in the UV limit, we know
\be
\sigma(\omega)\to \sigma_0\ \ \ \ (\omega\to \infty), \label{crt}
\ee
where $\sigma_0$ is a constant and is given by $T_p$ in our holographic calculation.
To maintain the damping condition, we define
\be
G_R(\omega)=-i\la J_x(\omega)J_x(0)\lb_R+i\sigma_0\omega,
\ee
and we apply the sum rule (\ref{sumr}). When $\sigma(\omega)$ is a smooth function, this leads to
the sum rule of the conductivity
\be
\int^\infty_0 d\omega \left(\mbox{Re}\,\sigma(\omega)-\sigma_0\right)=0,
\ee
because we can set $G_R(0)=0$ (or equally there is no pole in $\mathrm{Im}\, \sigma(\omega)$).

When there is a delta function Drude peak, if we define $n$ by $\sigma(\omega)\to \f{in}{\omega}$ in
the limit $\omega\to 0$, we obtain the following modified sum rule
\be
\int^\infty_{0+} d\omega \left(\mbox{Re}\,\sigma(\omega)-\sigma_0\right)=-\f{\pi}{2}n,
\ee
where notice that the integral does not include the contribution from the delta functional
part.

It will be instructive to compare the above with the standard one in non-relativistic
condensed matter systems (see e.g.\cite{Mahan}). The latter is given by
\be
\int^\infty_0 d\omega\, \mbox{Re}\,\sigma(\omega)=2\pi n_0>0,
\ee
where $n_0$ is proportional to the electron density.
This is obtained by assuming the behavior of plasma oscillation
\be
\sigma(\omega)\to \f{in_0}{\omega},\ \ \ \ \ \ (\omega\to \infty). \label{nonr}
\ee
The difference of the UV behavior between (\ref{crt}) and (\ref{nonr}) is crucial to have two
different sum rules.

\end{document}